\documentclass[amssymb,aps,pre,reprint,groupedaddress,
amsmath,amsfonts,amsthm]{revtex4-1}
\usepackage{graphicx}
\usepackage{float}
\usepackage{verbatim}

\begin{document}

\title{Hypostatic jammed packings of frictionless nonspherical particles}

\author{Kyle VanderWerf$^1$, Weiwei Jin$^{3,4}$, Mark D. Shattuck$^2$, Corey S. O'Hern$^{4,1,5,6}$}
\affiliation{$^1$ Department of Physics, Yale University, New Haven, Connecticut 06520, USA \\
$^2$ Benjamin Levich Institute and Physics Department, \\
The City College of New York, New York, New York 10031, USA \\
$^3$ Department of Mechanics and Engineering Science, Peking University, Beijing 100871, China \\
$^4$ Department of Mechanical Engineering \& Materials Science, \\
Yale University, New Haven, Connecticut 06520, USA \\
$^5$ Department of Applied Physics, Yale University, New Haven, Connecticut 06520, USA \\
$^6$ Graduate Program in Computational Biology and Bioinformatics, \\
Yale University, New Haven, Connecticut 06520, USA}

\date{\today}

\begin{abstract}
We perform computational studies of static packings of a variety of
nonspherical particles including circulo-lines, circulo-polygons,
ellipses, asymmetric dimers, dumbbells, and others to determine which
shapes form packings with fewer contacts than degrees of freedom
(hypostatic packings) and which have equal numbers of contacts and degrees
of freedom (isostatic packings), and to understand why hypostatic packings of
nonspherical particles can be mechanically stable despite having fewer
contacts than that predicted from na\"ive constraint counting. To
generate highly accurate force- and torque-balanced packings of
circulo-lines and -polygons, we developed an interparticle potential
that gives continuous forces and torques as a function of the particle
coordinates. We show that the packing fraction and coordination number
at jamming onset obey a master-like form for all of the nonspherical
particle packings we studied when plotted versus the particle
asphericity ${\cal A}$, which is proportional to the ratio of the
squared perimeter to the area of the particle. Further, the eigenvalue
spectra of the dynamical matrix for packings of different particle
shapes collapse when plotted at the same ${\cal A}$. For hypostatic
packings of nonspherical particles, we verify that the number of
``quartic'' modes along which the potential energy increases as the
fourth power of the perturbation amplitude matches the number of
missing contacts relative to the isostatic value.  We show that the
fourth derivatives of the total potential energy in the directions of
the quartic modes remain nonzero as the pressure of the packings is
decreased to zero.  In addition, we calculate the principal curvatures
of the inequality constraints for each contact in circulo-line
packings and identify specific types of contacts with inequality
constraints that possess convex curvature.  These contacts can
constrain multiple degrees of freedom and allow hypostatic packings of
nonspherical particles to be mechanically stable.
\end{abstract}


\maketitle


\section{Introduction}
\label{intro}

\begin{figure}
\centering
\includegraphics[width=0.5\textwidth]{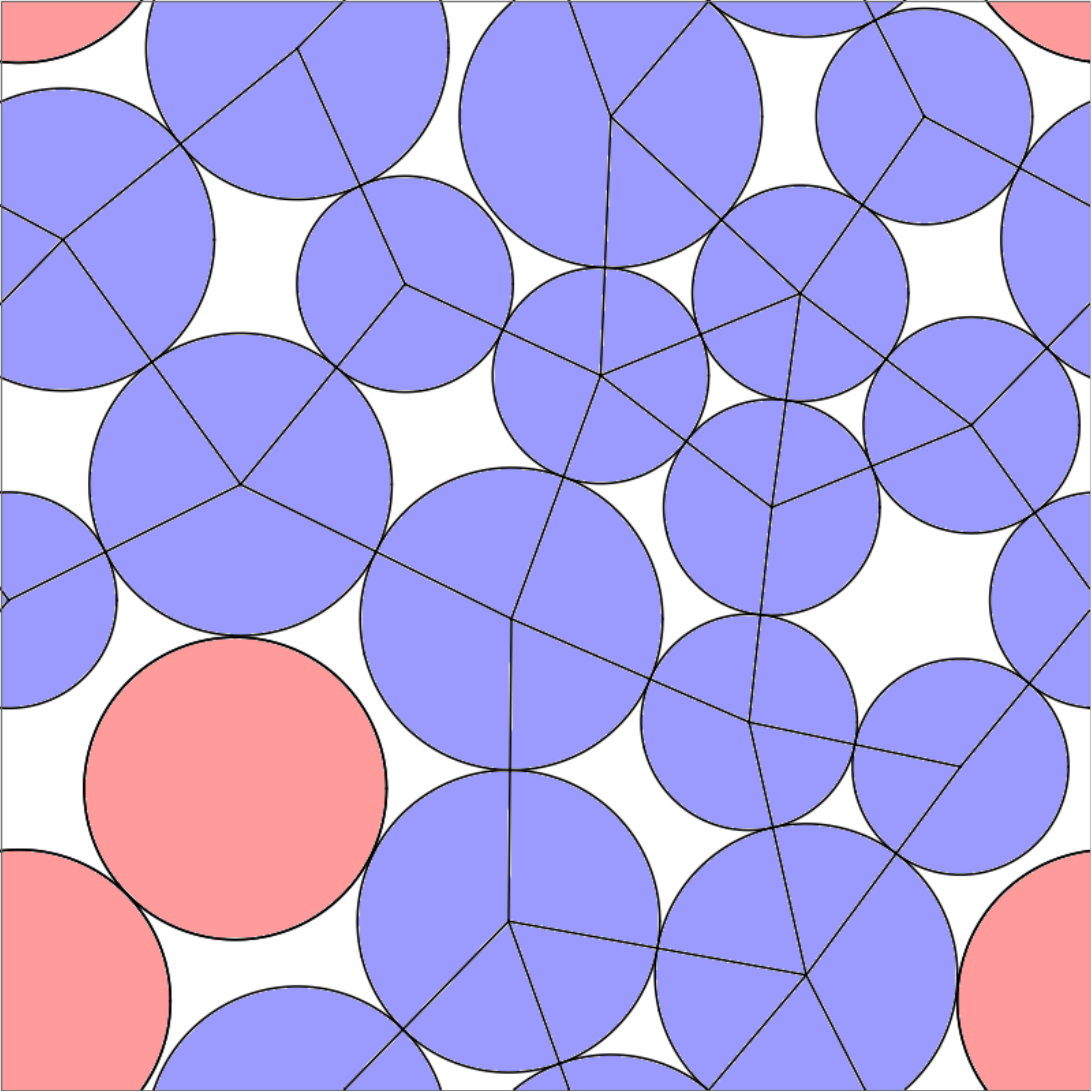}
\caption{An isostatic packing of $N=18$ bidisperse disks ($9$ large 
and $9$ small with diameter ratio $r=1.4$) in a square box with 
periodic boundary conditions at jamming onset $\phi_J = 0.835$. 
Blue disks form the force-bearing backbone, and red disks are 
``rattler'' disks that possess fewer than $3$ contacts.  This 
isostatic packing possesses $N_c = N_c^0 = 2(N-N_r) - 1 = 31$ 
contacts, where $N_c^0$ is the isostatic number of contacts.}
\label{fig:isodisk}
\end{figure}

There have been a significant number of computational studies aimed at
elucidating the jamming transition in static packings of frictionless
spherical particles~\cite{ohernmonodis,liu,vanhecke}.  Key findings
from these studies include: i) sphere packings at
jamming onset at packing fraction $\phi_J$ are isostatic (where the
number of contacts matches the number of degrees of freedom, as shown
in Fig.~\ref{fig:isodisk}), ii) the coordination number, shear modulus,
and other structural and mechanical quantities display power-law
scaling as a function of the system's pressure $P$ as packings are
compressed above jamming onset at $P=0$, and iii) the density of
vibrational modes develops a plateau at low frequencies $\omega$ that
extends toward $\omega \rightarrow 0$ as the system approaches jamming
onset. Many of these results are robust with respect to changes in the
particle size polydispersity and different forms for the purely
repulsive interparticle potential.

Most studies of jamming to date have been performed on
packings of disks in 2D or spheres in 3D.  More recently, both
computational and experimental studies have begun focusing on packings
of nonspherical shapes, such as
ellipsoids~\cite{donevellipse,hypo-ellipse-early,schreckdynmat,zeravcicellipse,ellipseexpt1,schaller,chaikin1,chaikin2},
spherocylinders~\cite{spherocyl,scylexpt,williams,meng,wouterse,wouterserod},
polyhedra~\cite{torquatopoly,chenpoly}, and composite
particles~\cite{dimerellipse,gaines,miskin,baule}. In particular,
there is a well-established set of results on packings of frictionless
ellipses (or ellipsoids in 3D).  In general, static packings of
frictionless ellipses are {\it hypostatic} with fewer contacts than
the number of degrees of freedom using na\"ive contact counting.  For
amorphous mechanically stable (MS) packings of disks, the coordination
number in the large-system limit is $z=2d_f$ (where $d_f=2$ is the
number of degrees of freedom per particle)~\cite{tkachenko}.  Thus,
one might expect that the coordination number for ellipses in 2D would
jump from $z=4$ to $z=6$ (with $d_f=3$) for any aspect ratio
$\alpha>1$. However, $z(\alpha)$ increases continuously from $4$ at
$\alpha=1$ and remains less than $6$ for all $\alpha$.  We have shown
that the number of missing contacts exactly matches the number of
``quartic'' eigenmodes from the dynamical matrix for which the
potential energy increases as the fourth power of the displacement for
perturbations along the corresponding
eigenmode~\cite{schreckdynmat}. In addition, the packing fraction at
jamming onset $\phi_J(\alpha)$ possesses a peak near $\alpha \approx
1.5$, and then decreases for increasing $\alpha$.

Are these results for ellipses similar to those for all other
nonspherical or elongated particle shapes?  Prior results for packings
of spherocylinders have shown that they are
hypostatic~\cite{wouterse}.  However, packings of composite particles
formed from collections of disks (2D)~\cite{dimerellipse,papanikolaou}
or spheres (3D)~\cite{gaines} are isostatic at jamming
onset. Unfortunately, very few studies explicitly check whether
hypostatic packings are mechanically stable.  The goal of this article
is to determine which particle shapes can form mechanically stable
({\it i.e.} jammed), hypostatic packings, identify a key shape
parameter that controls the forms of the coordination number $z$ and
packing fraction $\phi_J$, and gain a fundamental understanding for
why hypostatic packings are mechanically stable, even at jamming onset
$P=0$.

To address these questions, we generate static packings using a
compression and decompression scheme coupled with energy minimization
for nine different convex particle shapes (ellipses, circulo-lines,
circulo-triangles, circulo-pentagons, circulo-octogons,
circulo-decagons~\cite{wang}, dimers~\cite{dimerellipse},
dumbbells~\cite{han}, and Reuleaux triangles~\cite{atkinson}) in 2D.
To study such a wide range of particle shapes, we developed a fully
continuous and differentiable interparticle potential for
circulo-lines and circulo-polygons, which allows us to generate
packings with extremely accurate force and torque balance at very low
pressure.  We find several important results.
First, we show that the jammed packing fraction for the nine particle
shapes collapses onto a master-like curve when plotted versus the
asphericity parameter ${\cal A} = p^2/4\pi a$, where $p$ is the
perimeter and $a$ is the area of the particle.  We also show that the
coordination number $z({\cal A})$ follows a master-like curve when
contacts between nearly parallel circulo-lines or nearly parallel
sides of circulo-polygons are treated properly. In addition, for
packings of circulo-lines, we calculate the fourth derivatives of the
total potential energy along the quartic modes of the dynamical
matrix~\cite{schreckdynmat} and show that the fourth derivatives are
nonzero as $P \rightarrow 0$, which proves that these hypostatic
packings are mechanically stable.  Finally, we calculate the principal
curvatures of the constraint surfaces in configuration space defined
by each contact to identify which types of contacts in packings of
circulo-lines allow them to be mechanically stable, while hypostatic.

This article is organized as follows. In Sec.~\ref{sec:methods}, we
describe the compression and decompression plus minimization method we
use to generate static packings of convex, nonspherical particles. In
Sec.~\ref{sec:hypo}, we present examples of static packings of several
different convex, nonspherical particle shapes to put forward a
conjecture concerning which nonspherical particle shapes form
hypostatic packings and which always form isostatic packings. In
Sec.~\ref{sec:bulk}, we show that the packing fraction at jamming
onset $\phi({\cal A})$ and coordination number $z({\cal A})$ display
master-like forms when plotted versus the particle asphericity ${\cal
  A}$, for nine different nonspherical particle shapes. In this
section, we also show results for the calculations of the fourth
derivatives of the total potential of the static packings in the
directions of dynamical matrix eigenmodes. Finally, in
Sec.~\ref{sec:diffgeom}, we calculate and analyze the principal
curvatures of the constraint surfaces given by the interparticle
contacts to understand the grain-scale mechanisms that enable
hypostatic packings to be mechanically stable. We also include three
Appendices. Appendix~\ref{app:potential} describes the development of
a continuous interparticle repulsive potential between circulo-lines
and circulo-polygons, which allows us to generate extremely accurate
force- and torque-balanced jammed packings near
zero pressure. Appendix~\ref{app:polyshape} describes how we generate
different circulo-polygon shapes at constant asphericity ${\cal
  A}$. Finally, in Appendix~\ref{app:dynmat}, we provide expressions
for the elements of the dynamical matrix for packings of
circulo-polygons.

\section{\label{sec:methods}Methods}

\begin{table*}
\caption{A list of the nine convex, nonspherical particle shapes studied in
this article, along with the ranges of aspect ratio $\alpha$ and 
asphericity ${\cal A}$ that we considered.}
\label{tab:shapes}
\begin{ruledtabular}
\begin{tabular}{ccc}
Particle Shape & $\alpha-1$ & ${\cal A}-1$ \\
\hline
Circulo-Line & $10^{-3}$ -- $4$ & $4.05\times10^{-7}$ -- $1.06$\\
Circulo-Triangle & -- & $10^{-6.5}$ -- $0.4$ \\
Circulo-Pentagon & -- & $10^{-6.5}$ -- $0.4$ \\
Circulo-Octagon & -- & $10^{-6.5}$ -- $0.4$ \\
Circulo-Decagon & -- & $10^{-6.5}$ -- $0.1$ \\
Dimer & $0.571$ & $0.349$ \\
Dumbbell & $2$ -- $4$ & $0.514$ -- $2.405$ \\
Reuleaux Triangle & -- & $0.114$ \\
Ellipse & $10^{-4}$ -- $0.9$ & $3.75\times10^{-9}$  -- $0.161$ \\
\end{tabular}
\end{ruledtabular}
\end{table*}

Using computer simulations, we generate static packings of
frictionless, nonspherical, convex particles in 2D. The particles are
nearly hard in the sense that we consider mechanically stable packings in the
zero-pressure limit. We study nine different particle shapes:
circulo-lines, circulo-triangles, circulo-pentagons, circulo-octagons,
circulo-decagons, Reuleaux triangles, ellipses, dumbbells, and
dimers. (See Table~\ref{tab:shapes}.) We focus on bidisperse mixtures
in which half of the particles are large and half are small to prevent
crystallization~\cite{oldmonodis,ohernmonodis}. The large particles
have areas that satisfy $a_L = 1.4^2 a_S$, where $a_{L,S}$ is the area
of the large and small particles, respectively. Both particles have
the same mass, $m$. We generated static packings at fixed asphericity
${\cal A}$ for the large and small particles over a wide range of
${\cal A}$.  We employ periodic boundary conditions in square domains
with edge length $L=1$ and system sizes that vary from $N=24$ to $480$
particles. Note that the term ``convex particle shapes'' stands for
``shapes whose accessible contact surface is nowhere locally
concave.'' Our studies include dimers (which possess two points on the
surface that are concave), circulo-lines (which contain regions of
zero curvature), ellipses, and other explicitly convex particles.

\begin{figure}
\centering
\includegraphics[width=0.5\textwidth]{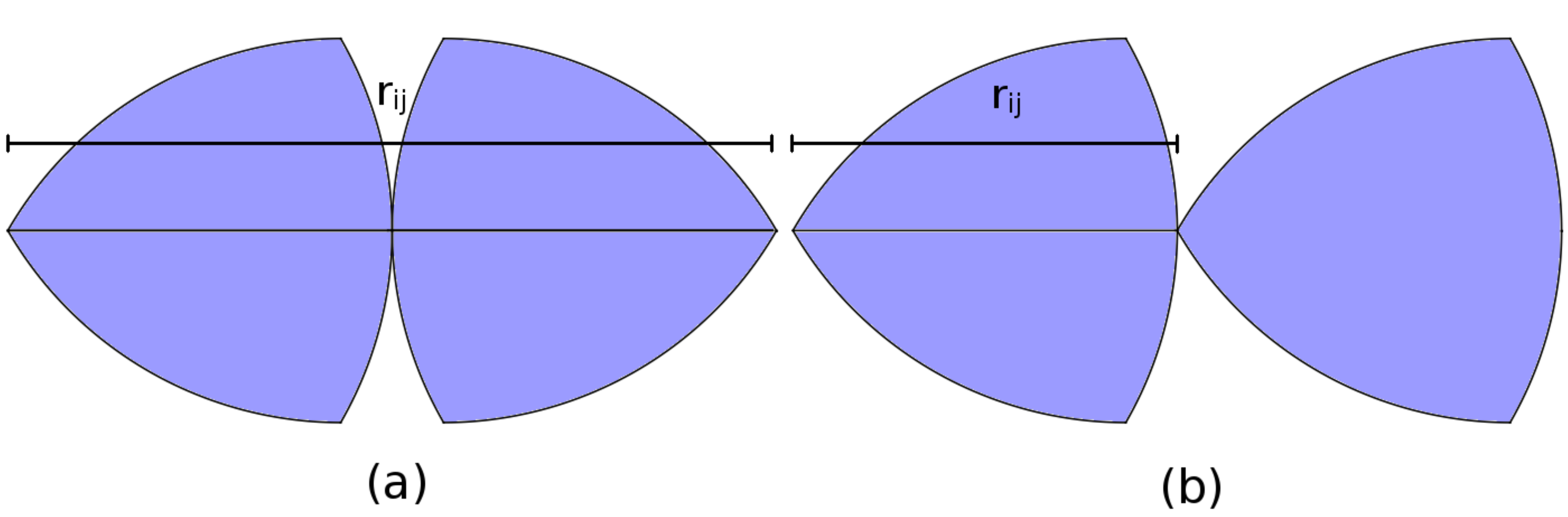}
\caption{Definition of the separation $r_{ij}$ between two
Reuleaux triangles when (a) the two circular
arcs are overlapping and (b) a vertex is overlapping one of the circular
arcs. In case (a), $r_{ij}$ is the distance between the
vertices at the centers of the corresponding arcs. In case (b),
$r_{ij}$ is the distance between the arc's central vertex and the
vertex overlapping the arc.}
\label{fig:reuleauxinteraction}
\end{figure}

We assume that particles $i$ and $j$ interact via the purely
repulsive, pairwise linear spring potential,
\begin{equation}
U(r_{ij}) = \frac{k}{2} \left(\sigma_{ij}- r_{ij} \right)^2
\Theta \left(\sigma_{ij}- r_{ij} \right),
\label{potential}
\end{equation}
where $k$ is the spring constant of the interaction and $\Theta(x)$ is
the Heaviside step function. Below, lengths, energies, and pressures
will be expressed in units of $L$, $kL^2$, and $k$, respectively. For
disks, $r_{ij}$ is the separation between the centers of disks $i$ and
$j$, and $\sigma_{ij}=R_i + R_j$ is the sum of the radii of disks $i$
and $j$.

For dimers, {\it i.e.} composite particles formed from two
circular monomers, $r_{ij}$ is the center-to-center separation between
each pair of interacting monomers, and $\sigma_{ij}$ is the sum of the
radii of those monomers. A Reuleaux triangle is a shape that is
constructed by joining three circular arcs of equal radius such that
their intersection points (vertices of the Reuleaux triangle) are the
centers of each circle. For this shape, we first identify whether two
arcs are overlapping or whether a vertex is overlapping an arc.  We
then set $r_{ij}$ in Eq.~\ref{potential} to be the distance between
the center points of the overlapping arcs (in the case of two
overlapping arcs) or the distance between the center point of the arc and
the vertex (in the case of a vertex overlapping an arc). We set
$\sigma_{ij}$ to be the sum of the radii of the two overlapping arcs,
or the radius of the single arc when a vertex is overlapping an arc. (See
Fig.~\ref{fig:reuleauxinteraction}.)

For ellipses, we take $r_{ij}$ to be the distance between the centers
of the particles, and $\sigma_{ij}$ to be the center-center distance
that would bring the particles exactly into contact at their current
orientations~\cite{schreckdynmat}. For dumbbell-shaped particles, we
have multiple possible cases for $r_{ij}$, depending on their
orientations. We calculate $r_{ij}$ either as the distance between
each pair of circular ends, or between each circular end and the other
particle's shaft, with $\sigma_{ij}$ chosen to be the sum of the
relevant radii in each case. The repulsive contact interactions between
circulo-lines and -polygons are calculated in a similar fashion to
dumbbells. However, because the regions of changing curvature in the
case of circulo-lines and -polygons are accessible, unlike in the
dumbbell case, additional constraints in the potential are necessary
to prevent discontinuities in the pairwise torques and forces. For a
thorough explanation of how we define a continuous, repulsive linear
spring potential between circulo-lines and polygons, see
Appendix~\ref{app:potential}.

To generate static packings, we successively compress and decompress
the system with each compression or decompression step followed by the
conjugate gradient method to minimize the total potential energy
$U=\sum_{i>j} U(r_{ij})$. We use a binary search algorithm to push the
system to a target pressure $P=P_0$.  If $P > P_0$, the system is
decompressed isotropically, and if $P < P_0$, the system is compressed
isotropically. Subsequently, we perform minimization of the
enthalpy~\cite{smith} $H=U+P_0 A$, where $A$ is the area of the
system, the pressure $P=-dU/dA$, and $P_0=10^{-9}$ is the target
pressure, with the particle positions and the box edge length as the
degrees of freedom. Using this algorithm, we achieve accurate force
and torque balance such that the squared forces $f^2_i$ and torques
$\tau_i^2$ on a given particle $i$ do not exceed $10^{-25}$.

After generating each static packing, we calculate its dynamical
matrix $M$, which is the Hessian matrix of second derivatives of the total 
potential energy $U$ with respect to the particle coordinates:
\begin{equation}
M_{ij} = \frac{\partial^2 U}{\partial \xi_i \partial \xi_j},
\end{equation}
where $\xi_i=x_i$, $y_i$, and $\theta_i$, $x_i$ and $y_i$ are the
coordinates of the geometric center of particle $i$, and $\theta_i$
characterizes the rotation angle of particle $i$. We then
calculate the $3N$ eigenvalues $\lambda_i$ of $M$, and the
corresponding eigenvectors ${\vec \lambda}_i$ with ${\vec
  \lambda}^2_i=1$. For more details on the calculation of the
dynamical matrix elements, see Appendix~\ref{app:dynmat}.

\section{Results}
\label{sec:results}

Our results are organized into three subsections. In
Sec.~\ref{sec:hypo}, we discuss which nonspherical particle shapes
give rise to hypostatic packings, and then propose specific criteria
that nonspherical particle shapes must satisfy to yield hypostatic
packings. In Sec.~\ref{sec:bulk}, we show the variation of the packing
fraction $\phi$ and coordination number $z$ at jamming onset with
particle asphericity ${\cal A}$ for packings of circulo-lines,
circulo-polygons, and ellipses. Finally, in Sec.~\ref{sec:diffgeom},
we calculate the principal curvatures of the inequality constraints in
configuration space arising from interparticle contacts for hypostatic
packings of circulo-lines to identify the specific types of contacts
that allow static packings to be hypostatic, yet mechanically stable.

\subsection{Nonspherical Particle Shapes that Give Rise to Hypostatic Packings}
\label{sec:hypo}

\begin{figure}
\centering
\includegraphics[width=0.5\textwidth]{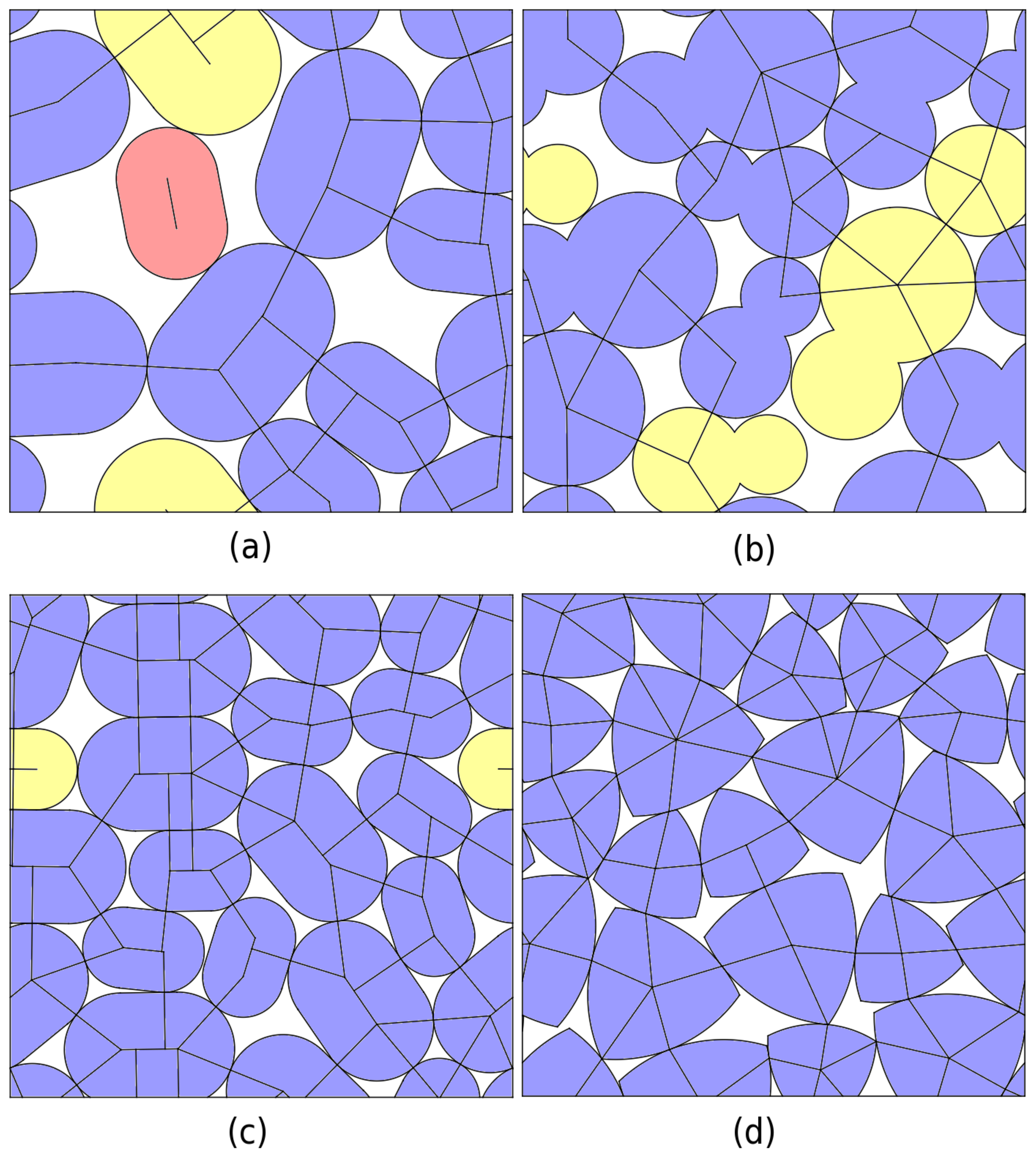}
\caption{Two isostatic ((a) and (b)) and two hypostatic packings ((c)
and (d)) of nonspherical particles. (a) This packing with $\phi =
0.782$ consists of $10$ non-rotating circulo-lines with asphericity
${\cal A} = 1.06$. The red particle is a rattler with two
unconstrained degrees of freedom, and the yellow particle is a
slider with one unconstrained degree of freedom. The blue and yellow particles
form an isostatic contact network with $N_c = 2N-4=16$ contacts, where we have
subtracted off $3$ additional contacts due to the rattler and slider
particles. (b) This packing with $\phi=0.828$ consists of $10$
asymmetric dimers. The two monomers on a given dimer have a diameter
ratio $r=1.4$ and the ratio of the lengths of large and small dimer
axes is $d=1.4$. The three yellow `rotator' particles each have 
one unconstrained rotational degree of freedom. The particles 
form an isostatic contact network with $N_c=3N-4=26$, where we have 
subtracted off $3$ additional contacts due to the rotator particles.  
(c) This packing with $\phi=0.892$ consists of $18$ rotating circulo-lines 
with asphericity ${\cal A} = 1.06$. The yellow particle is a slider 
with $1$ unconstrained degree of freedom. If the particles 
formed an isostatic contact network, it would possess $N_c = 3N-2=52$ 
contacts.  However, we find $N_c =46$. (d) This packing with $\phi = 0.874$
consists of $18$
Reuleaux triangles. If the system were isostatic, $N_c = 3N-1=53$, however, 
we find $N_c = 43$.}
\label{fig:all4packings}
\end{figure}

In this section, we discuss results for the contact number of static
packings containing a variety of nonspherical particle shapes. Based
on these results, we propose that frictionless convex particles will
form hypostatic packings if both of the following two criteria are
satisfied: (i) the particle has one or more nontrivial rotational degrees of
freedom, and (ii) the particle cannot be defined as a union of a
finite number of disks without changing its accessible
contact surface. Below, we show several examples of systems that
satisfy and do not satisfy these criteria.

First, disks do not satisfy (i) or (ii), and hence our conjecture
predicts that disks will form isostatic, not hypostatic, packings.
Next, we consider packings of circulo-lines that are prevented from
rotating, and thus the particle's orientation remains the same over the
course of the packing simulations. (See Fig.~\ref{fig:all4packings}
(a).) These particles obey criterion (ii), as a circulo-line can only
be expressed as an {\it infinite} union of disks, but fail to meet
criterion (i). Hence, the above conjecture predicts that these
particles will form isostatic, not hypostatic packings. We also
generated packings of bidisperse asymmetric dimers
(Fig.~\ref{fig:all4packings} (b)). These particles meet criterion (i),
since we allow them to rotate, but fail to meet criterion (ii), since
dimers are made up of a union of two disks. Thus, our conjecture
predicts that these particles will form isostatic, not hypostatic
packings, as shown in Fig.~\ref{fig:all4packings} (b).

Finally, we generated packings of rotating circulo-lines, as well as
Reuleaux triangles, examples of which are pictured in
Fig.~\ref{fig:all4packings} (c) and (d), respectively. Both particles
meet criterion (i), since they are allowed to rotate. Circulo-lines
meet criterion (ii) as stated earlier. Reuleaux triangles also meet
criterion (ii). Despite being comprised of a finite number of circular
arcs, it is impossible to define them as a finite number of complete
disks. Therefore, since both particle shapes meet both criteria, our
conjecture predicts that they will form hypostatic, not isostatic
packings. Ellipses also meet criteria (i) and (ii) and form hypostatic
packings~\cite{schreckdynmat}.

\begin{figure}
\centering
\includegraphics[width=0.5\textwidth]{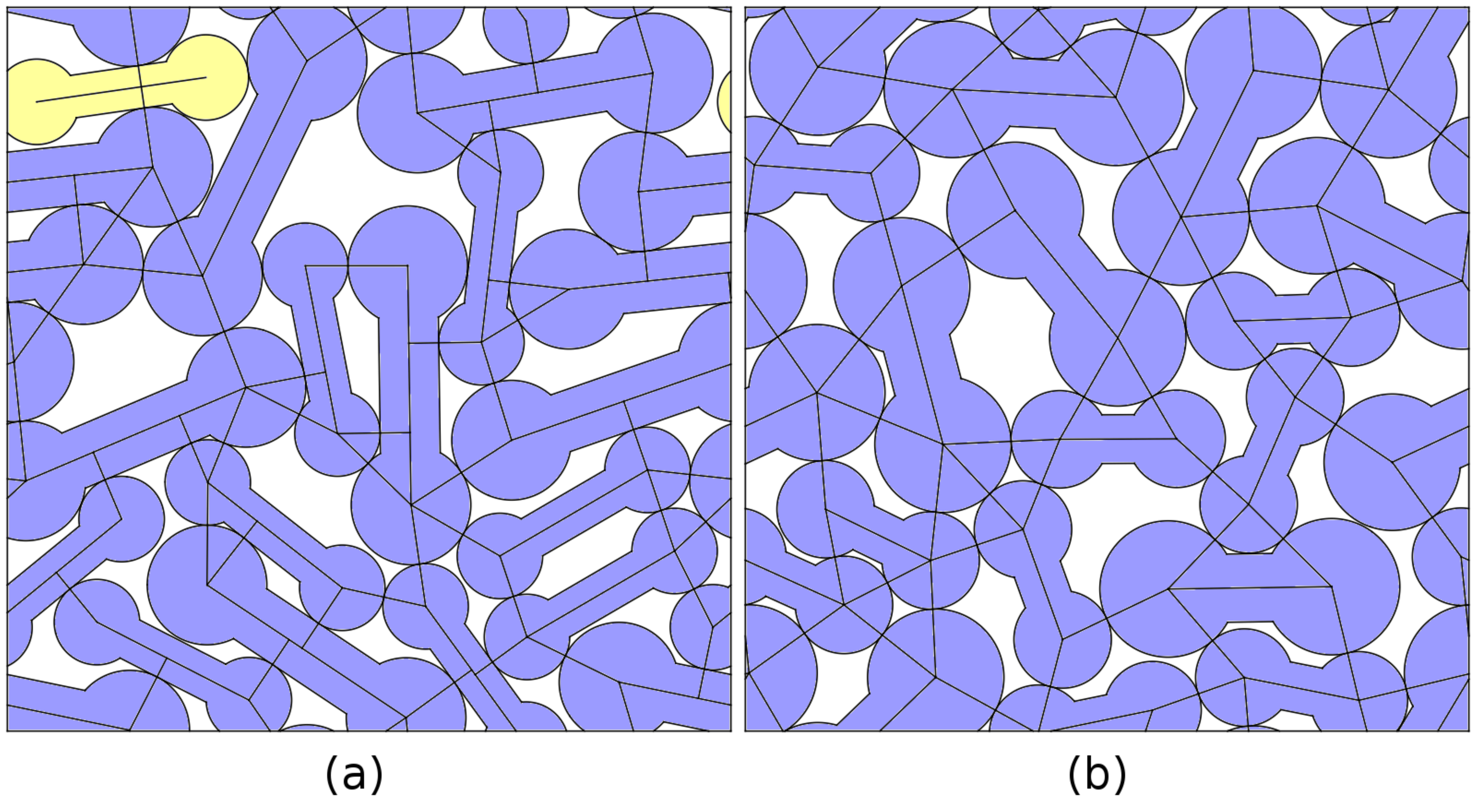}
\caption{(a) A hypostatic and (b) an isostatic packing of $18$ dumbbells,
with packing fractions $\phi = 0.769$ and $0.806$, respectively. In
(a) and (b), the shaft half-width is equal to half of the radius of
the end disks. In (a) the length of the shaft is $4$ times the disk
radius, whereas in (b), the shaft length is $2.4$ times the disk
radius. In (a), the yellow particle has one unconstrained degree of freedom. 
Thus, if the particles formed an isostatic contact network, $N_c 
= 3N-2=52$. The packing is hypostatic with $N_c =51$. In (b), the packing 
has no particles with unconstrained degrees of freedom.  This packing 
is isostatic with $N_c = 3N-1=53$.}
\label{fig:dumbbell}
\end{figure}

The importance of specifying ``accessible contact surface'' in
criterion (ii) can be demonstrated by the two packings of dumbbells in
Fig.~\ref{fig:dumbbell}. In both cases, the particles are allowed to
rotate, so criterion (i) is satisfied. The packing in (a) also
satisfies criterion (ii) because the shaft is part of the accessible
contact surface of the constituent particles, and the shaft cannot be
defined as a finite union of disks. Thus, we expect hypostatic
packings for the dumbbells in Fig.~\ref{fig:dumbbell} (a). In
contrast, in Fig.~\ref{fig:dumbbell} (b), the shaft is not part of the
accessible contact surface, because it is too short to allow the end
disks of other particles to come into contact with it. Thus, the
particles in Fig.~\ref{fig:dumbbell} (b) do not satisfy criterion
(ii), because the accessible contact surface is a union of two
disks. We expect packings generated using the dumbbells in
Fig.~\ref{fig:dumbbell} (b) to be isostatic.

\subsection{Packing Fraction, Coordination Number, and Eigenvalues of 
the Dynamical Matrix}
\label{sec:bulk}

In this section, we describe studies of the packing fraction and
coordination number of packings of nonspherical particles at jamming
onset as a function of the particle asphericity ${\cal A}$. We also
calculate the eigenvalues of the dynamical matrix for packings of
circulo-lines and circulo-polygons and show the eigenvalue spectrum as
a function of decreasing pressure. We find that hypostatic packings
possess a band of eigenvalues, {\it i.e.} the `quartic modes', for
which the energy increases as the fourth power in amplitude when we
perturb the system along their eigendirections.  These quartic modes
are not observed in isostatic packings. We further show that the
fourth derivative of the total potential energy in the direction of
these quartic modes does not vanish at zero pressure, proving that
packings possessing quartic modes are mechanically stable, despite
being hypostatic.

\begin{figure}
\centering
\includegraphics[width=0.5\textwidth]{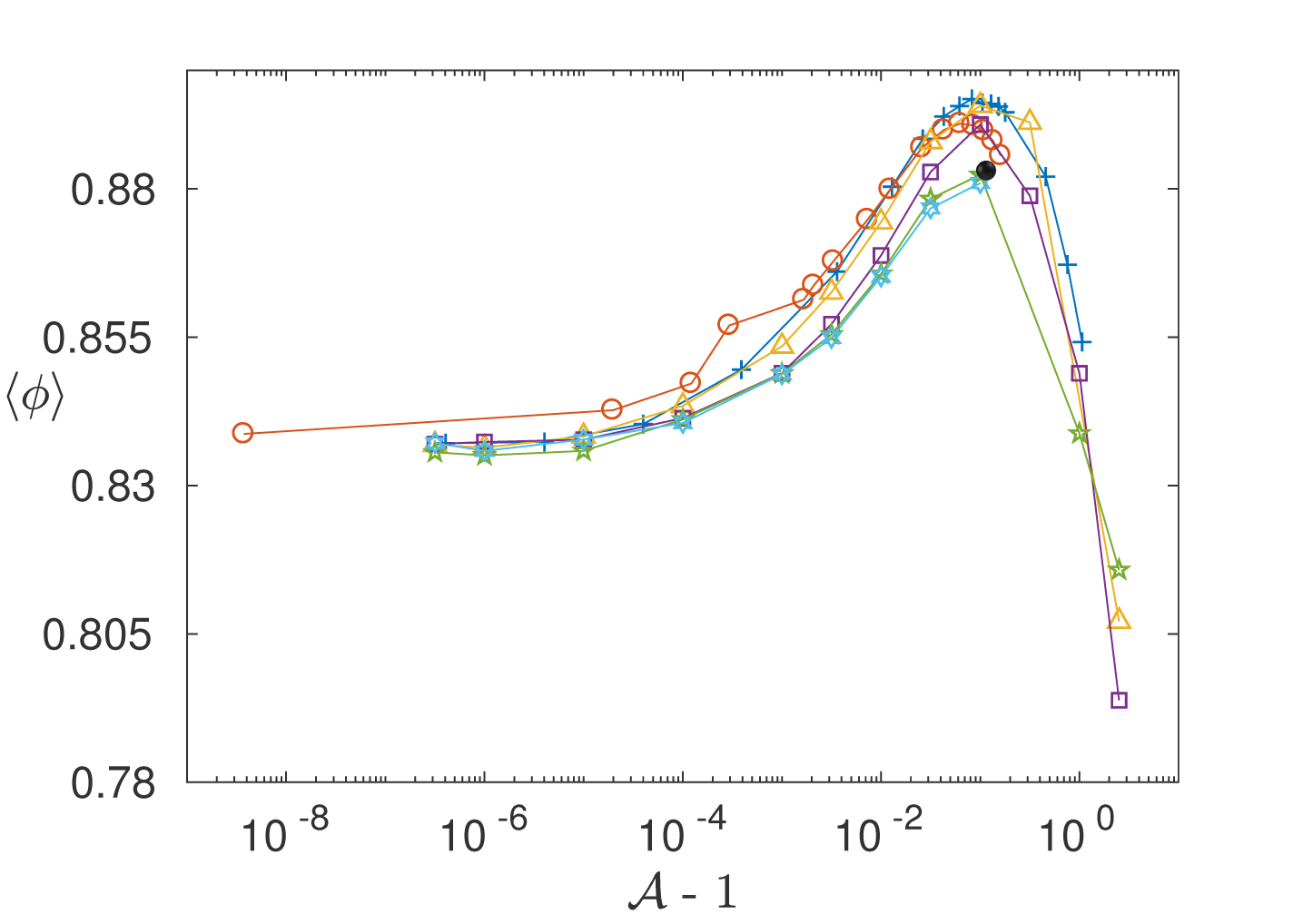}
\caption{Packing fraction at jamming onset $\langle \phi \rangle$ (averaged
over $50$ packings with random initial conditions) plotted versus asphericity ${\cal A}$ for a
variety of nonspherical shapes: ellipses (circles), circulo-lines (plus signs), circulo-triangles
(triangles), circulo-pentagons (squares), circulo-octagons
(five-pointed stars), circulo-decagons (six-pointed stars), and
Reuleaux triangles (filled black circle).  The packings include 
$N=100$ particles, except for the ellipse packings, which contain $N=480$
particles.}
\label{fig:pfracvsar}
\end{figure}

In Fig.~\ref{fig:pfracvsar}, we plot the average packing fraction at
jamming onset versus ${\cal A}$ for all of the nonspherical particles
we considered.  The data for $\langle \phi \rangle$ nearly collapses
onto a master curve, which tends to $\langle \phi \rangle \approx
0.84$ for small ${\cal A}-1$, as found for packings of bidisperse
disks~\cite{corey-algo}, forms a peak near ${\cal A}-1 \approx
10^{-1}$, and decreases strongly for ${\cal A}-1 > 10^{-1}$.  This
result suggests that the asphericity can serve as common descriptor of
the structural and mechanical properties of packings of nonspherical
particles, {\it i.e.} jammed packings with similar ${\cal A}$ will
possess similar properties.

\begin{figure}
\centering
\includegraphics[width=0.5\textwidth]{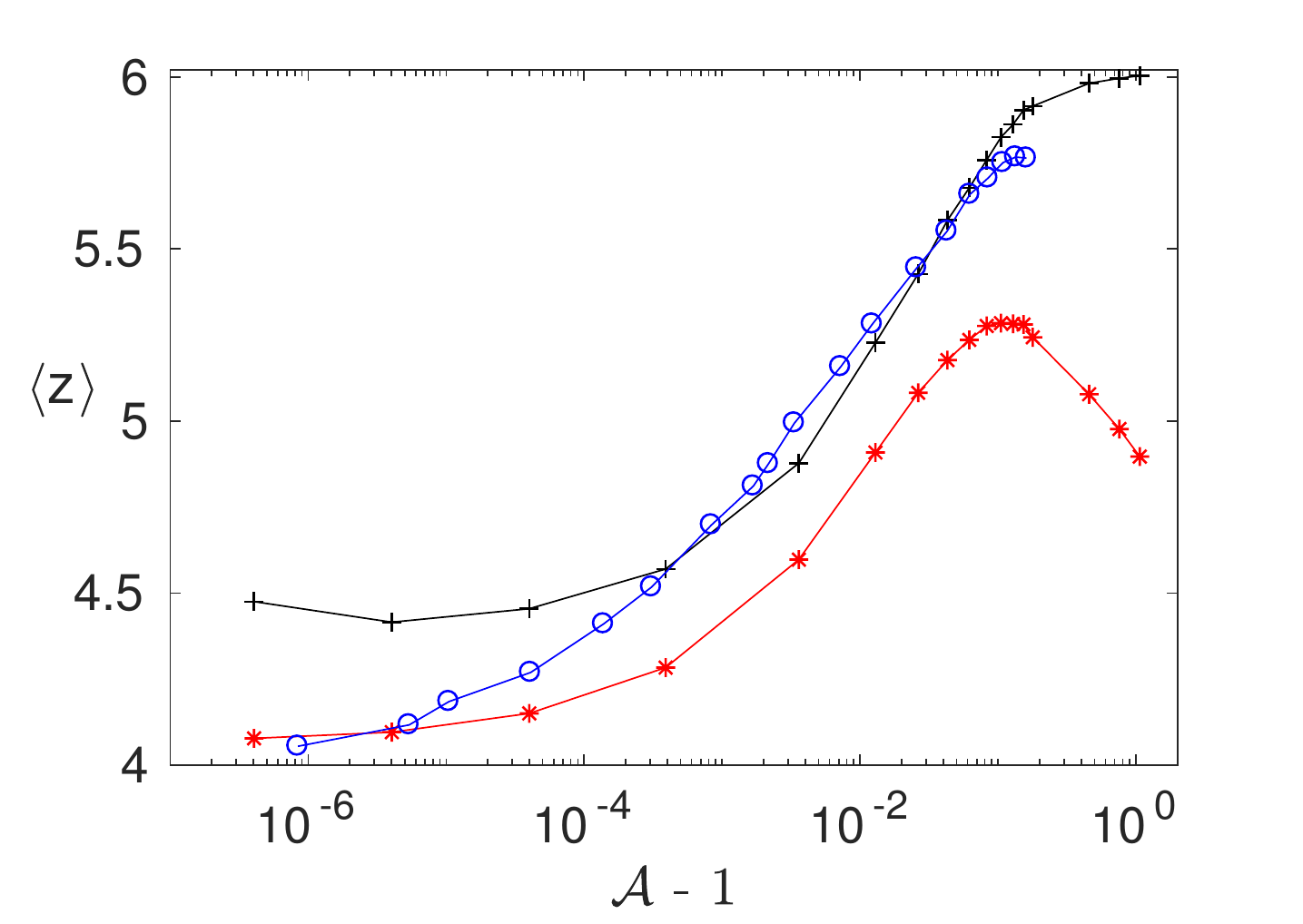}
\caption{Coordination number $\langle z \rangle$ 
(defined in Eq.~\ref{z_adjusted}) as a function of asphericity
${\cal A}-1$ for ellipses (circles) and circulo-lines, counting 
a contact between nearly parallel circulo-lines as one contact (asterisks) 
or two contacts (plus signs).}
\label{fig:coordvsar}
\end{figure}

In Fig.~\ref{fig:coordvsar}, we plot the average coordination number
\begin{equation}
\langle z \rangle = \frac{2(N_c+1)}{N-N_r-N_s/3},
\label{z_adjusted}
\end{equation}
where $N_c$ is the number of contacts in the packing.  The $+1$ in the
factor of $N_c+1$ is included to account for the $-1$ in the
expression for the number of contacts $N_c = N_c^0 = 3N-1$ in
isostatic packings of nonspherical particles in 2D, where $N_c^0$ 
is the isostatic number of contacts. $N_r$ is the number of
rattler particles that have unconstrained translational and rotational
degrees of freedom. $N_s$ is the number of `slider' particles with a
single unconstrained translational degree of freedom. An example of a
slider particle is the yellow particle in the packing of circulo-lines
in Fig.~\ref{fig:all4packings} (c), which can translate along its long
axis without energy cost.  Defining the coordination number as in
Eq.~\ref{z_adjusted} ensures that an isostatic packing of
circulo-lines, circulo-polygons, or other nonspherical particles will
have $\langle z\rangle=6$.  If $\langle z\rangle < 6$, the packing is hypostatic. 

In Fig.~\ref{fig:coordvsar}, we show the coordination number $\langle
z \rangle $ in Eq.~\ref{z_adjusted} versus ${\cal A}-1$ for packings
of ellipses and circulo-lines for two ways of defining a contact
between two nearly parallel circulo-lines. At low asphericities, where
the particle shape approaches a disk, a nearly parallel contact is
only able to apply a small torque to the two contacting particles,
making it unlikely to constrain a rotational degree of freedom in
addition to a translational degree of freedom. Thus, at low
asphericities, nearly parallel contacts should only be counted as a
single constraint.  In Fig.~\ref{fig:coordvsar}, we show that $\langle
z \rangle$ for ellipses and circulo-lines (counting nearly parallel
contacts once) both approach $4$ in the limit ${\cal A} -1$ tends to
zero.

In contrast, at large asphericities, nearly parallel contacts between
two circulo-lines prevent the particles from rotating and translating
(in a direction perpendicular to their shafts). Thus, for large ${\cal
  A}-1$, nearly parallel contacts should be counted as two
constraints.  In Fig.~\ref{fig:coordvsar}, we show that the
coordination number $\langle z \rangle$ for packings of circulo-lines
approaches $6$ in the large ${\cal A}-1$ limit when nearly parallel
contacts are counted twice.  These results suggest that we must
interpolate between counting parallel contacts once at low
asphericities, and counting them twice as the asphericity increases.

One way to resolve the question of whether to count a nearly parallel
contact between nonspherical particles as one or two constraints is
to calculate the dynamical matrix (all second derivatives of the total
potential energy with respect to the particle coordinates) of the
static packings, and examine the spectrum of the dynamical matrix
eigenvalues, which in the harmonic approximation give the vibrational
frequencies of the packing~\cite{tanguy}. For details on the calculation of the
entries of the dynamical matrix for circulo-lines and
-polygons, see Appendix~\ref{app:dynmat}.

\begin{figure}
\centering
\includegraphics[width=0.5\textwidth]{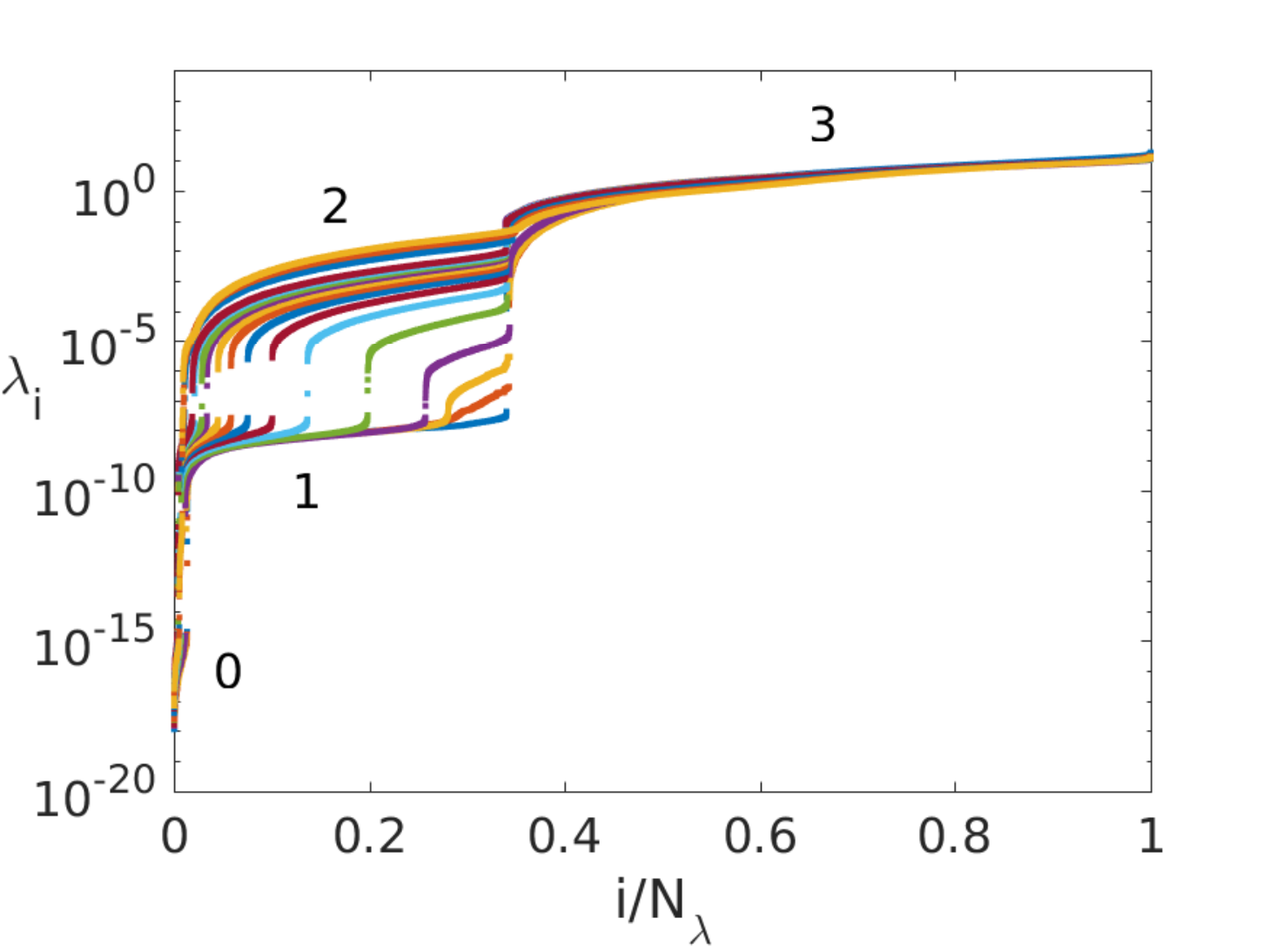}
\caption{The eigenvalues $\lambda_i$ of the dynamical matrix sorted from smallest 
to largest for static packings of $N=100$ 
circulo-lines at the $17$ different asphericities ${\cal A}-1$ shown 
in Fig.~\ref{fig:coordvsar} ranging from 
$\approx 10^{-6}$ 
to $1$ and decreasing from top to bottom. $N_{\lambda}=15000$ is the total 
number of eigenvalues in all of the packings at a given asphericity. For each asphericity (different 
colors), we show spectra for $50$ separate packings. We label four 
distinct regions of the eigenvalue spectra, $0$-$3$.  Region $0$ 
($\lambda_i \lesssim 10^{-14}$) corresponds to unconstrained degrees of 
freedom (such as overall translations due to periodic boundary conditions, rattler and slider
particles), 
region $1$ ($10^{-14} \lesssim \lambda_i \lesssim 4 \times 10^{-8}$) 
corresponds to ``quartic 
modes,'' whose number is determined by the number of missing contacts 
relative to the isostatic contact number, region $2$ corresponds to eigenmodes with predominantly rotational 
motion, and region $3$ corresponds to eigenmodes with predominantly 
translational motion.}
\label{fig:eigenbranchesscyl}
\end{figure}

In Fig.~\ref{fig:eigenbranchesscyl}, we show the eigenvalue spectrum
(sorted from smallest to largest) for static packings of circulo-lines
over a wide range of aspect ratios ${\cal A}-1$ from $\approx 10^{-6}$
to $1$ (decreasing from top to bottom).  As found in
Ref.~\cite{hypo-ellipse-early} for ellipse packings, the eigenvalue
spectra for packings of circulo-lines possess several distinct
regions. Region $0$ ($\lambda_i \lesssim 10^{-14}$, which is
set by numerical precision) corresponds to unconstrained
degrees of freedom, such as overall translations from periodic boundary 
conditions, and rattler and slider
particles. Region $1$ ($10^{-14} \lesssim \lambda_i \lesssim 4 \times
10^{-8}$) corresponds to ``quartic modes,'' whose number is determined
by the number of missing contacts relative to the isostatic contact
number.  For the asphericities we consider, regions $2$ and $3$
correspond to eigenmodes with predominantly rotational and
translational motion, respectively.

If we focus on all but the three smallest asphericities ({\it i.e.} the
three rightmost curves in Fig.~\ref{fig:eigenbranchesscyl}), we can
define a cutoff value $\lambda_c$ that clearly separates regions $1$
and $2$. For packings of $N=100$ circulo-lines at pressure
$P_0=10^{-9}$, $\lambda_{c} \approx 4 \times 10^{-8}$. For asphericities
where $\lambda_c$ distinguishes regions $1$ and $2$, the number of
contacts in packings of circulo-lines satisfies $N_c = N^{0}_c - N_1$,
where $N_1$ is the number of eigenvalues in region $1$.  A key
observation is that defining the number of contacts in this way for
intermediate and high asphericities is the same as if $N_c$ is
determined by the number of particle contacts, with nearly
parallel contacts counted twice. For asphericities where the
difference between regions $1$ and $2$ is more ambiguous, we still use
$\lambda_c$ to determine whether a given eigenvalue belongs to region $1$ or
$2$. For the lowest asphericities, we find that defining $N_c = N^0_c-N_1$
corresponds to counting one constraint for each nearly parallel
contact.

\begin{figure}
\centering
\includegraphics[width=0.5\textwidth]{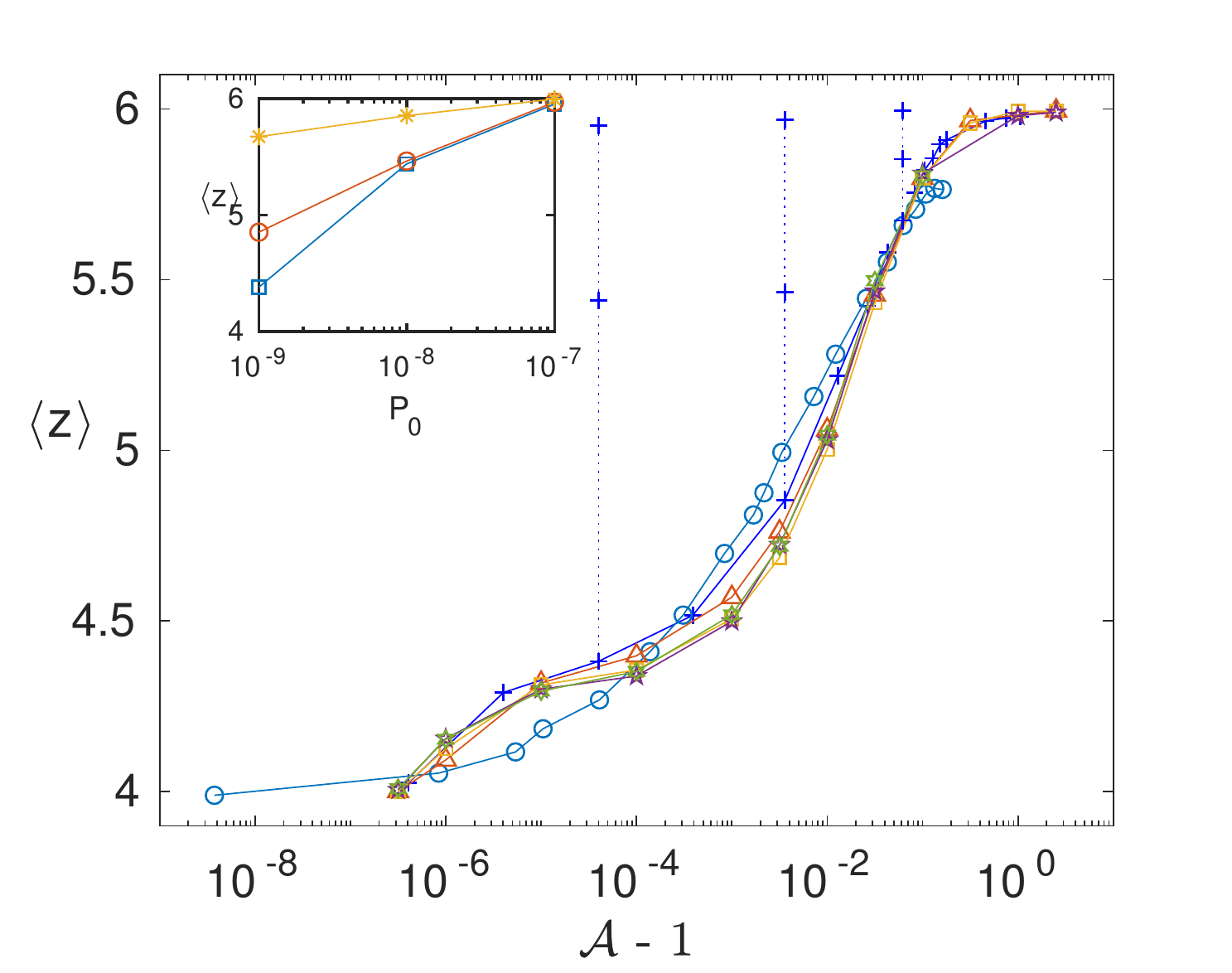}
\caption{The average coordination number $\langle z\rangle$ 
(Eq.~\ref{z_adjusted}) is plotted 
versus asphericity ${\cal A}-1$ for packings of $N=100$ circulo-lines 
(plus signs), -triangles
(triangles), -pentagons (squares), -octagons
(five-pointed stars), and -decagons (six-pointed stars). This data  
is compared to $\langle z \rangle$ for packings of $N=480$ 
ellipses (circles). For packings of circulo-lines and -polygons, 
we define $\langle z \rangle$ using $N_c = N^0_c - N_1$.  
In the inset, we show $\langle z \rangle$ as a function of
pressure $P_0$ for packings of circulo-lines for 
${\cal A}-1=4\times 10^{-5}$ (squares),
$3.6\times 10^{-3}$ (circles), and $6.2\times
10^{-2}$ (asterisks). $\langle z\rangle$ decreases more rapidly
with $P$ at lower asphericities. All data points in the inset 
are also plotted in the main figure (plus signs connected by 
dotted lines).}
\label{fig:eigencoordvsar}
\end{figure}

In Fig.~\ref{fig:eigencoordvsar}, we plot the average coordination
number $\langle z\rangle$ from Eq.~\ref{z_adjusted} using $N_c =
N^0_c-N_1$ versus ${\cal A}-1$ for $N=100$ packings of circulo-lines
and circulo-polygons. At low asphericities ${\cal A}-1$, the
coordination number for packings of circulo-lines and
circulo-polygons, as well as ellipses, approaches $\langle z \rangle =
4$, which is expected for bidisperse disk packings.  At large
asphericities, $\langle z \rangle = 6$ for packings of circulo-lines
and circulo-polygons as expected for isostatic packings with $2$
translational and $1$ rotational degree of freedom per
particle. $\langle z \rangle$ for ellipse packings plateaus for large
${\cal A}-1$. However, the current data suggests that the plateau
value is less than $6$, indicating that ellipse packings are hypostatic 
for all ${\cal A}-1$.  

An interesting feature in $\langle z \rangle({\cal A})$ for static
packings of circulo-lines and -polygons is the plateau in $\langle z
\rangle$ that occurs near ${\cal A}-1 \approx 10^{-5}$ in
Fig.~\ref{fig:eigencoordvsar}. Our results suggest that the plateau is
likely an artifact of the small, but nonzero pressure of the static
packings. If the particles are overcompressed, even slightly,
nearly parallel contacts will be able to exert larger torques than
they would at zero pressure, which causes more eigenvalues to be above
the eigenvalue threshold $\lambda_c$, and contacts to be counted as
two constraints instead of one. Thus, as we decrease the pressure to
zero, we expect to count fewer of these nearly parallel contacts as
two constraints and the plateau in $\langle z\rangle$ near ${\cal A}-1
\approx 10^{-5}$ will decrease.  As ${\cal A}-1$ decreases below
$10^{-5}$, the effects from overcompression are less important, 
and the nearly parallel contacts are only counted once.

\begin{figure}
\centering
\includegraphics[width=0.5\textwidth]{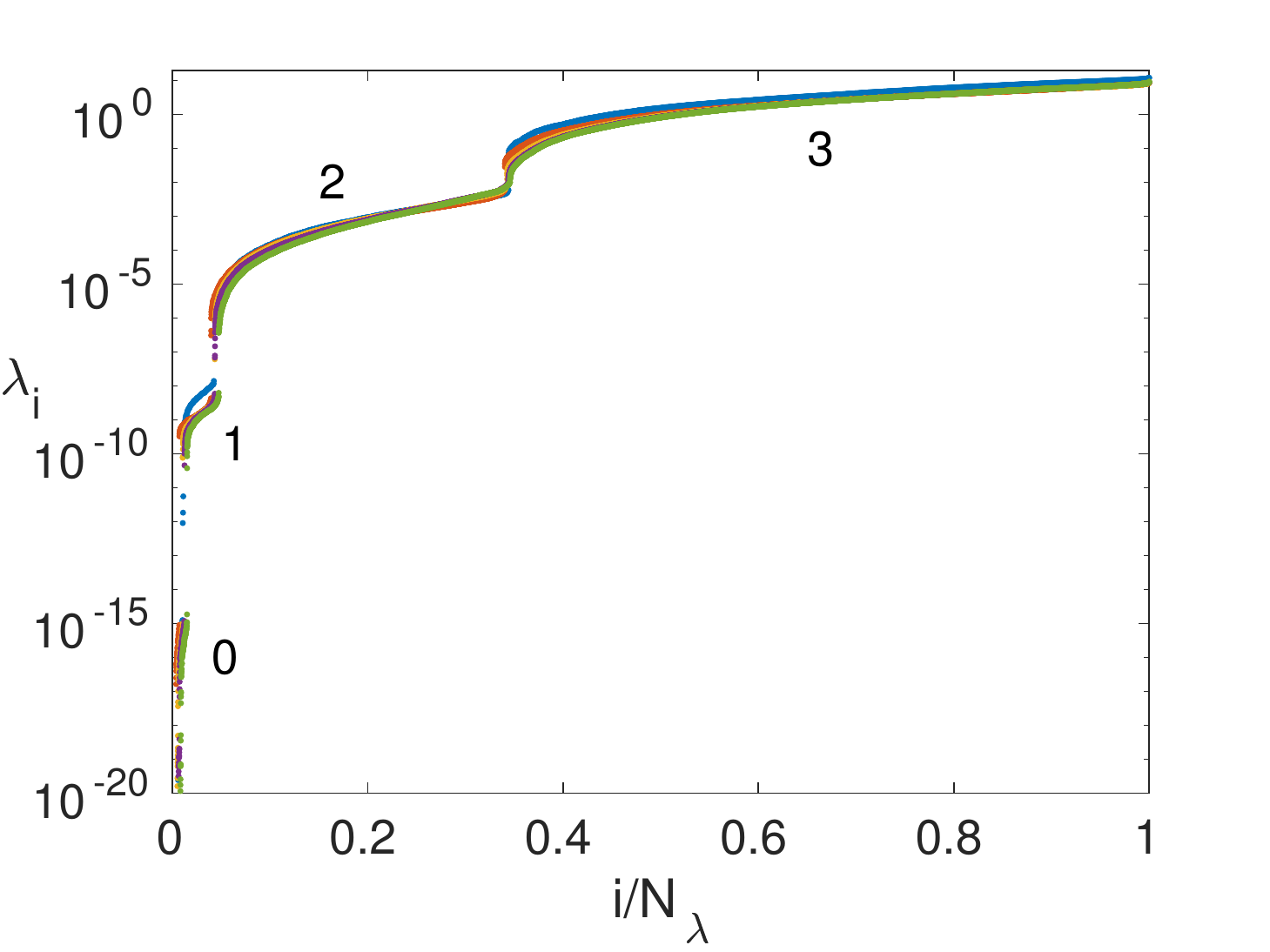}
\caption{The eigenvalues $i$ of the dynamical matrix sorted from smallest 
to largest for static packings of $N=100$ circulo-lines (blue), 
-triangles (red),
-pentagons (yellow), -octagons (purple), and
-decagons (green) at asphericity ${\cal A} =1.1$. $N_{\lambda}=15000$ is the total 
number of eigenvalues in all of the packings of a given shape. For each particle shape, we 
show spectra for $50$ separate packings.  Regions $0$-$3$ are 
defined the same way as in Fig.~\ref{fig:eigenbranchesscyl}.}
\label{fig:eigenbranchespolygon}
\end{figure}

In Fig.~\ref{fig:eigenbranchespolygon}, we plot the eigenvalues
$\lambda_i$ sorted from smallest to largest for static packings of
five different particle shapes (circulo-lines, -triangles, -pentagons,
-octagons, and decagons) at the same asphericity, ${\cal A}=1.1$. We
find that the eigenvalue spectra for all of these shapes are nearly
identical. This behavior differs markedly from that in
Fig.~\ref{fig:eigenbranchesscyl}, where we show the eigenvalue spectra
for packings with the same particle shape (circulo-lines), but at different values of
the asphericity. Circulo-polygons with $n$ sides possess $2n-3$
parameters that specify their shape (not counting uniform scaling of
lengths). Our results suggest that asphericity is a key parameter in
determining the structure, geometry, and physical properties of
hypostatic packings.

\begin{figure}
\centering
\includegraphics[width=0.5\textwidth]{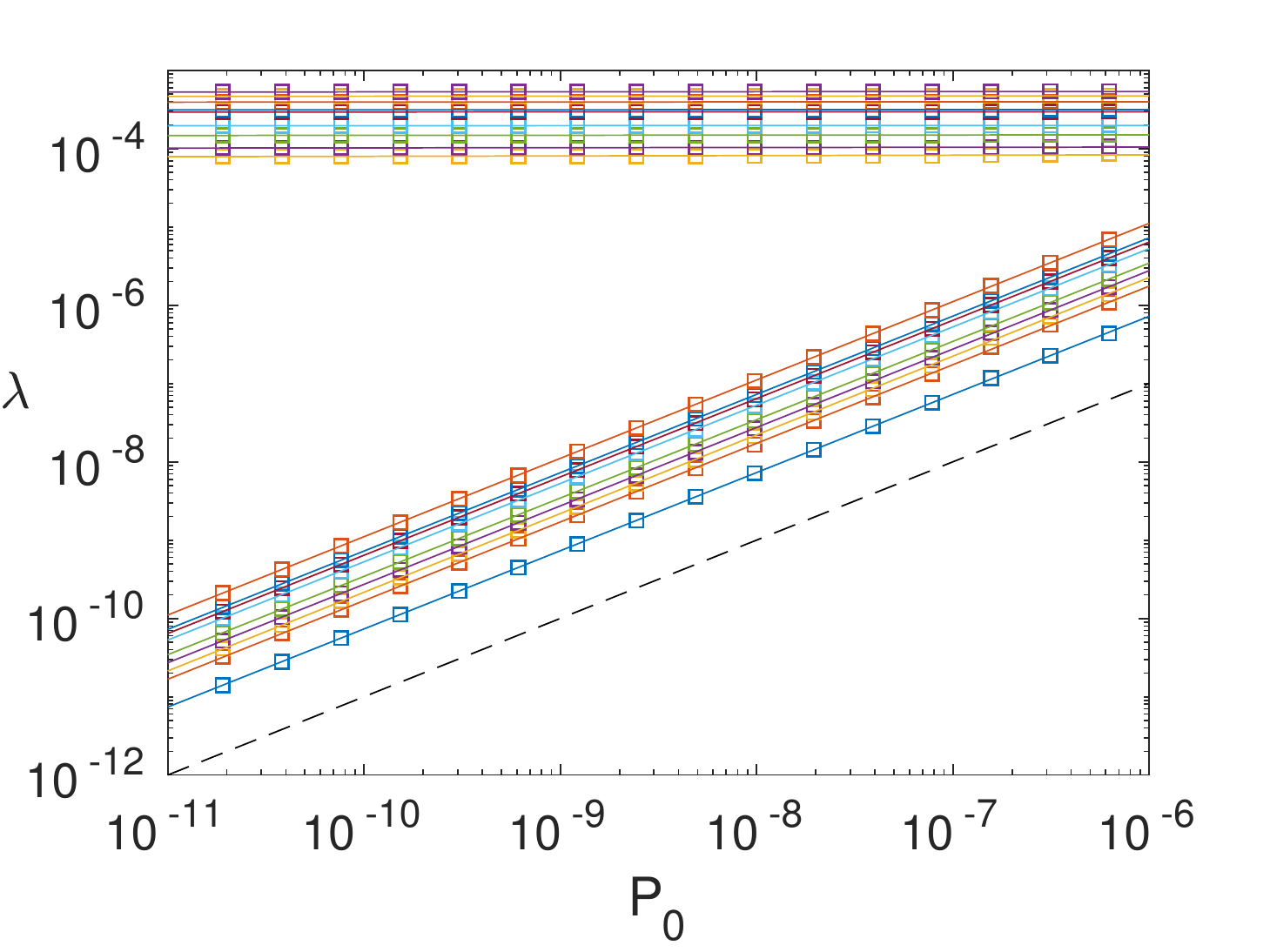}
\caption{Nine eigenvalues $\lambda$ of the dynamical matrix from regions $1$ 
(bottom) and $2$ (top) plotted versus pressure $P_0$ for a static 
packing of $N=32$ circulo-lines with asphericity ${\cal A}=1.03$. The 
dashed line has slope $1$.}
\label{fig:eigenpress} 
\end{figure}

The reason why the eigenvalues in region $1$ ({\it c.f.}
Fig.~\ref{fig:eigenbranchesscyl}) are referred to as ``quartic modes''
is that, for perturbations along the corresponding eigenvectors, the
total potential energy scales quartically with the amplitude of the
perturbation, rather than quadratically, as one would expect for
mechanically stable
packings~\cite{hypo-ellipse-early,schreckdynmat}. In
Fig.~\ref{fig:eigenpress}, we plot eigenvalues from regions $1$ and
$2$ as a function of pressure $P_0$ for a static packing of $N=32$
circulo-lines at asphericity ${\cal A}=1.03$. The eigenvalues from
region $2$ are independent of pressure, whereas the eigenvalues from
region $1$ scale linearly with pressure. Thus, for packings of
circulo-lines and other particle shapes that yield hypostatic
packings, the eigenvalues corresponding to the quartic modes are zero
at jamming onset ($P_0=0$). This result agrees with prior studies of
hypostatic packings of ellipses and ellipsoids~\cite{schreckdynmat}.

\begin{figure}
\centering
\includegraphics[width=0.5\textwidth]{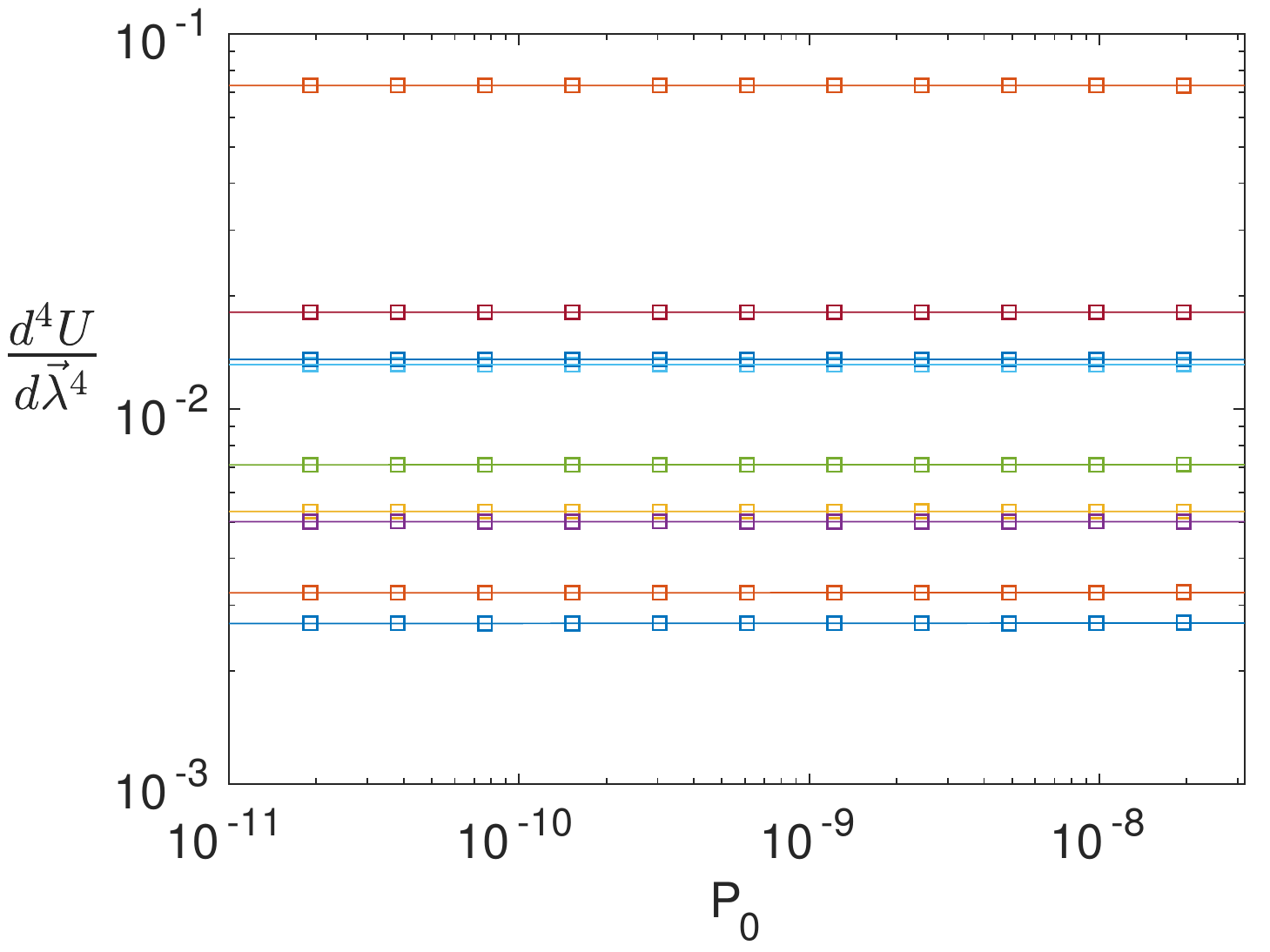}
\caption{For the same packing as in Fig.~\ref{fig:eigenpress}, we plot the
fourth derivative of the total potential energy $d^4U/d{\vec \lambda}^4$ 
in the direction of the
nine eigenmodes in region $1$ in the bottom of Fig.~\ref{fig:eigenpress} 
as a function of pressure $P_0$. All of the fourth derivatives for the 
region $1$ eigenmodes are independent of pressure.}
\label{fig:fourthpress}
\end{figure}

Perturbations along the quartic modes are constrained to fourth
order. In Fig.~\ref{fig:fourthpress}, we show the fourth derivatives
of the total potential energy $d^4U/d{{\vec \lambda}}^4$ in the directions of
the nine eigenmodes in region $1$ (that are depicted near the
bottom of Fig.~\ref{fig:eigenpress}).  We find that the fourth derivatives along
eigenmodes in region $1$ do not depend on pressure, and thus remain 
nonzero at zero pressure. These findings
demonstrate that hypostatic packings are fully constrained at zero
pressure---in some directions by quadratic potentials and in other
directions by quartic potentials.

\subsection{Convex versus concave constraints}
\label{sec:diffgeom}

Why are hypostatic packings of circulo-lines and other nonspherical
particles mechanically stable when they possess fewer contacts than
the isostatic number, $N_c < N_c^0$?  We have already shown that the
number of missing contacts $N_c^0-N_c$ matches the number of quartic
modes along which the energy increases quartically, not quadratically,
with the perturbation amplitude.  In the other $N_c$ eigendirections
of the dynamical matrix, the energy increases quadratically with the
perturbation amplitude.  As a result, there are no directions in
configuration space for which these hypostatic packings can be
perturbed without energy cost, and thus they are mechanically stable.

To more fully address the question of how hypostatic packings of
nonspherical particles can be mechanically stable, we consider the
so-called ``feasible region'' of configuration space near each static
packing for packing fractions slightly below jamming
onset~\cite{donevellipse}. The feasible region near a given static
packing includes all configurations for which there are no particle
overlaps. The boundaries of this region are determined by all of the
interparticle contacts, each of which corresponds to an inequality
among the particle coordinates specifying when pairs of particles do not
overlap. Points in configuration space that satisfy all of the
inequalities are inside the feasible region. For mechanically stable
packings, as the packing fraction is increased, the feasible region
shrinks and becomes bounded and compact, preventing particle
rearrangements that would allow the system to transition to a different 
packing. A static packing is mechanically stable if the
feasible region of accessible configurations shrinks to a single point
at jamming onset.

The number of constraints required to bound the feasible region
depends on the curvature of the inequality constraints in
configuration space, {\it i.e.} whether the constraints are concave or
convex~\cite{donevellipse}.  The inequality constraints that arise in disk packings are
always concave. In particular, in disk packings, the curvature of each
constraint is equal to minus the reciprocal of the sum of the radii of
the two disks in contact. As a result, the number of contacts required
to bound the feasible region for a mechanically stable packing of $N$
disks is $2N+1$ (minus $2$ from overall translations in periodic
boundary conditions).  Thus, {\it hypostatic} packings of nonspherical
particles must possess contacts that give rise to bounding surfaces
with convex curvature, which allows packings to be mechanically stable
with fewer than the isostatic number of contacts.

\begin{figure}
\centering
\includegraphics[width=0.5\textwidth]{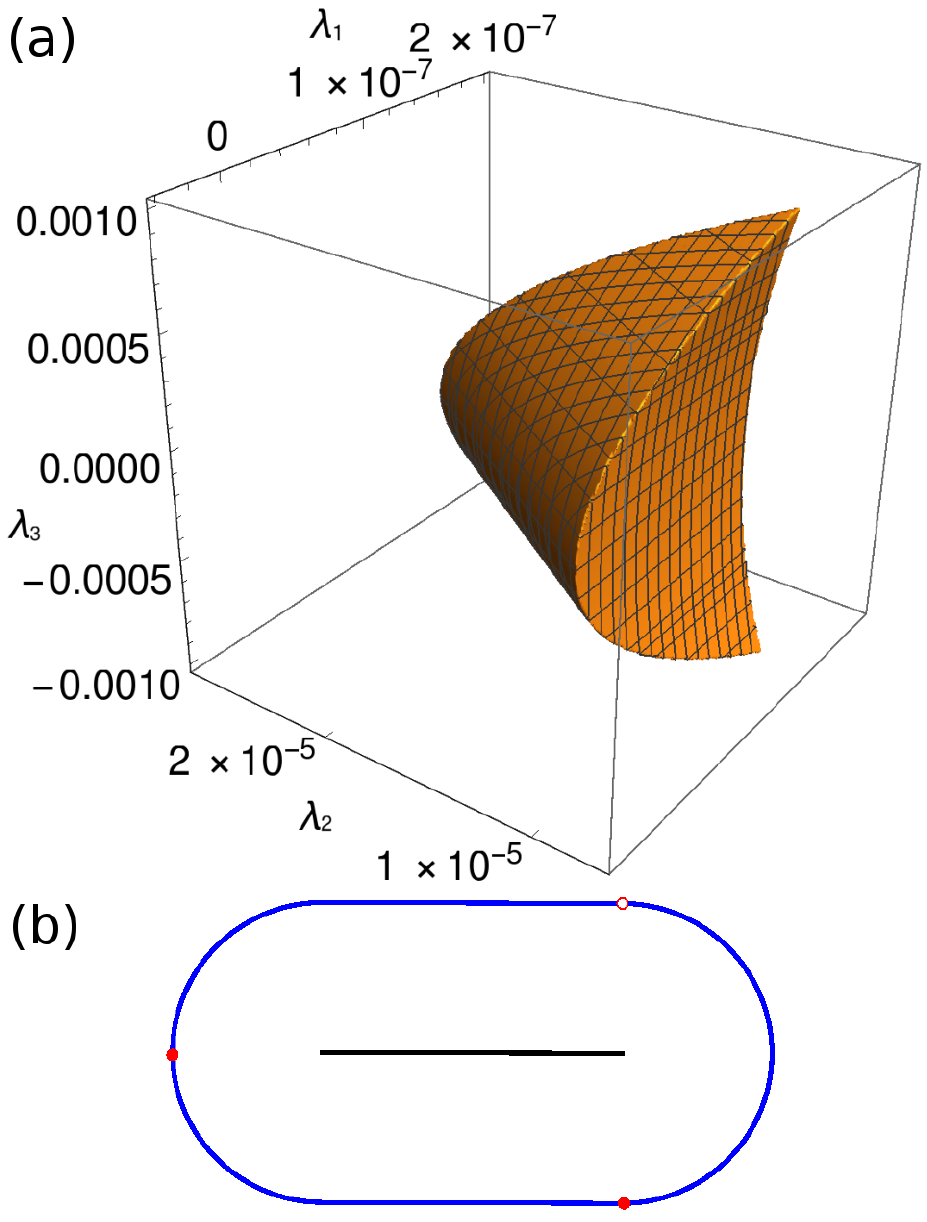}
\caption{(a) Depiction of the feasible region of configurations that
do not possess interparticle overlaps for a single circulo-line
surrounded by three fixed points as shown in panel (b). This
system is generated by fixing three points in space, initializing a
circulo-line between the three points, and growing the interior
circulo-line until it reaches force and torque balance while in
contact with the three points. After finding the stable
configuration, we decrease the diameter of the interior circulo-line by 
$10^{-7}$. The
extent of the feasible region is shown using coordinates along the
the three eigendirections (${\vec \lambda}_1$, ${\vec \lambda}_2$, and ${\vec
\lambda}_3$) of the dynamical matrix for the interior circulo-line.  
The top contact
(open circle) in (b), which is positioned along the shaft of the
circulo-line, provides the constraint with convex curvature.}
\label{fig:constrained}
\end{figure}

In Fig.~\ref{fig:constrained}, we show a simple configuration
involving a circulo-line that gives rise to a convex constraint. We
consider three points at fixed positions. These points represent less
strict constraints than contacts with other circulo-lines, and thus,
if these three points can constrain a circulo-line, three contacting
circulo-lines will constrain an interior circulo-line as well.  We
initialize a circulo-line at several locations between the three points, and
then increase the size of the interior circulo-line until it is
constrained by the three points.  After the circulo-line is constrained,
we shrink its diameter by $10^{-7}$ so
that it no longer overlaps the bounding points. The feasible region of
the slightly undercompressed circulo-line is shown in
Fig.~\ref{fig:constrained} (a).

For an isostatic system, four contacts are required to constrain a
circulo-line. However, we find configurations in which a circulo-line
is constrained by only three contacts. Fig.~\ref{fig:constrained} (a)
illustrates the reason that only three contacts are necessary: one of
the contacts (open circle on the top shaft) gives rise to a constraint
with convex curvature in configuration space. In contrast, the other
two contacts (filled circles), which are on the end caps of the
circulo-line, give rise to constraints with concave curvature. This
example suggests that only certain types of contacts between
circulo-lines generate constraints with convex curvature, and thus the
number of contacts required for mechanical stability is less than the
isostatic number when these types of contacts are present.

\begin{figure}
\centering
\includegraphics[width=0.5\textwidth]{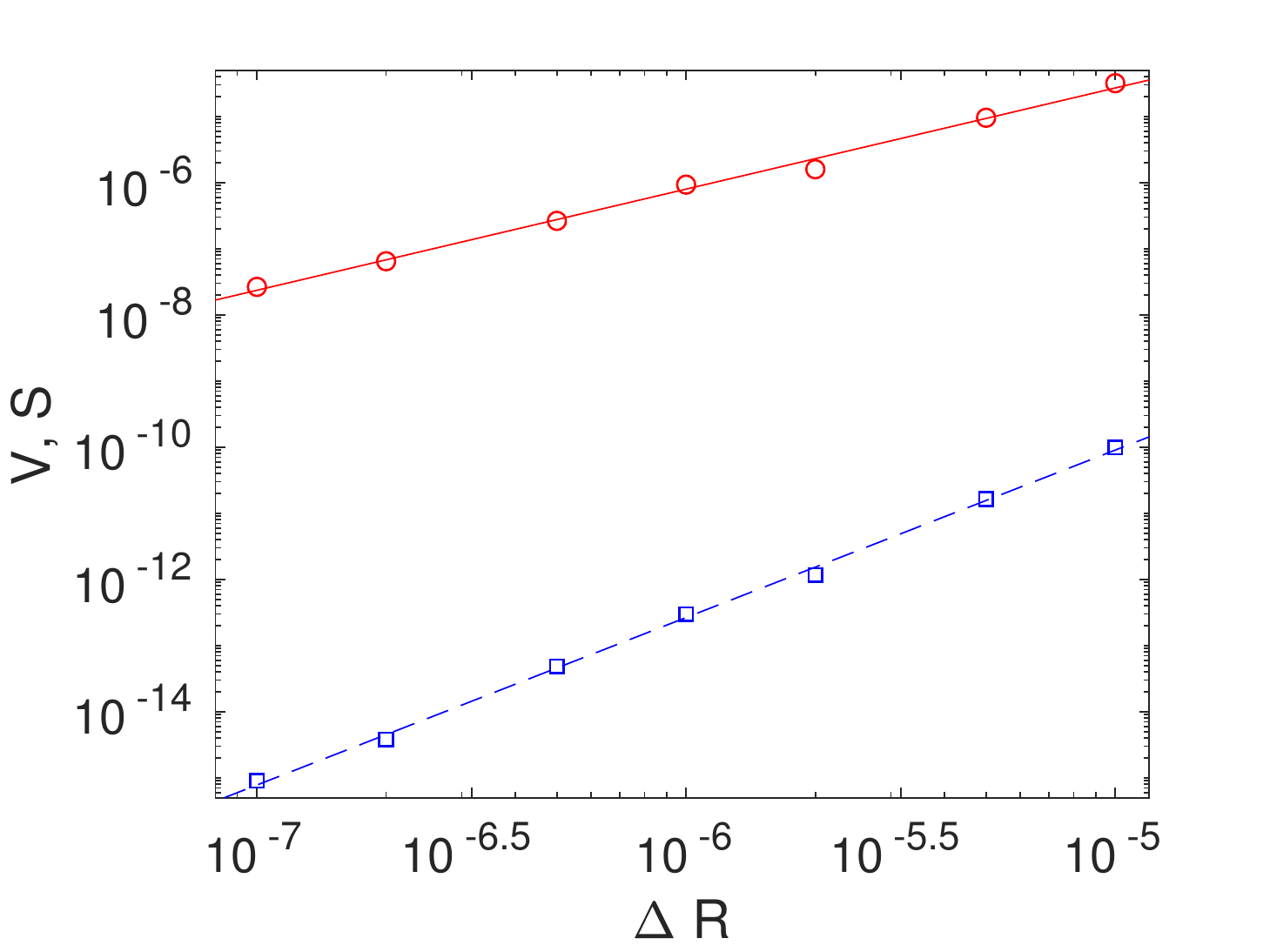}
\caption{The volume $V$ (squares) and surface area $S$ (circles) of the
feasible region in Fig.~\ref{fig:constrained} (a) plotted as a function of
the decrease in radius $\Delta R = R_j-R$, where $R_j$ is the radius at which 
the system is fully constrained. The solid and dashed lines
have slopes $2.5$ and $1.5$, respectively.}
\label{fig:volsadelr}
\end{figure}

To verify that the circulo-line ``packing'' in
Fig.~\ref{fig:constrained} (b) is mechanically stable, we numerically calculated
the volume $V$ and surface area $S$ of the feasible region as a
function of the degree of undercompression, $\Delta R=R_j - R$, where
$R_j$ is the radius of the interior circulo-line at which the system
is jammed. In Fig.~\ref{fig:volsadelr}, we show that both $V$ and
$S$ display power-law scaling with $\Delta R$, emphasizing
that the feasible region for hypostatic packings shrinks to a point,
and thus these packings are mechanically stable.

\begin{table}
\caption{The principal curvatures ($\langle \kappa_1 \rangle$ and
$\langle \kappa_2 \rangle$) for the different types of contacts that
can occur between two circulo-lines in a $N=24$ packing of
bidisperse circulo-lines with asphericity ${\cal A} = 1.04$,
averaged over all contacts in the packing of that type. Note that
for $\kappa_2$ for an end particle in a parallel contact, the
magnitude of the curvature was averaged, rather than the signed
curvature, because, unlike any of the other curvatures, this one
fluctuated between positive and negative. This type of averaging is
denoted using the $\pm$ symbol.}
\label{tab:example}
\begin{ruledtabular}
\begin{tabular}{ccc}
Contact Type & $\langle\kappa_1\rangle$ & $\langle\kappa_2\rangle$ \\
\hline
Parallel (End Particle) & $0$ & $(\pm) 7.93\times10^{-9}$\\
Parallel (Middle Particle) & $-1.29$ & $0.774$ \\
End-Middle (End Particle) & $-0.0261$ & $0$ \\
End-Middle (Middle Particle) & $-1.29$ & $0.774$ \\
End-End & $-5.69$ & $-0.0246$ \\
\end{tabular}
\end{ruledtabular}
\end{table}

To further investigate the effect of convex and concave constraints on a hypostatic jammed packing, we measured the curvatures of the
inequality constraints for each contact in a static packing with
$N=24$ bidisperse circulo-lines with asphericity ${\cal A}=1.04$. We classified
the contacts into five types as defined in
Appendix~\ref{app:potential}.  Parallel contacts can involve the
shaft of one circulo-line (middle) and the end cap of another
(end). This arrangement gives rise to two types of contacts, one for
the circulo-line with a contact on its end and another for the
circulo-line with a contact on its middle.  Similarly, the shaft
(middle) of one circulo-line can be in contact with the end cap (end) of
another, but the long axes are not parallel.  This arrangement again
gives rise to two types of contacts, one for the circulo-line with a
contact on its end and another for the circulo-line with a contact on its
middle.  In addition, the ends of two circulo-lines can be in contact.  

The average curvatures of the bounding surfaces for each contact type
in a static packing of $N=24$ bidisperse circulo-lines are compiled in
Table~\ref{tab:example}. (We find similar average values for other
$N=24$ packings of bidisperse circulo-lines.) From this data, we can
draw several conclusions about the contribution of each type of
contact to the stability of circulo-line packings. First, end-end
contacts yield concave constraints in configuration space, and thus on
their own do not give rise to mechanically stable hypostatic
packings. In contrast, end-middle contacts have a positive principal
curvature for the circulo-line whose middle is in contact, and thus
serve to stabilize hypostatic packings. Parallel contacts also possess
a positive curvature associated with the circulo-line whose middle is
in contact.  However, note that the concave curvature for
circulo-lines whose end is in \emph{parallel} contact is much smaller
than the concave curvature of the end circulo-line for end-middle
contacts. This means that for circulo-lines with end contacts, the
parallel contacts are more ``stabilizing'' than the end-middle
contacts, and therefore they are more frequent in mechanically
stable hypostatic circulo-line packings than other end-middle
contacts.

\begin{figure}
\centering
\includegraphics[width=0.5\textwidth]{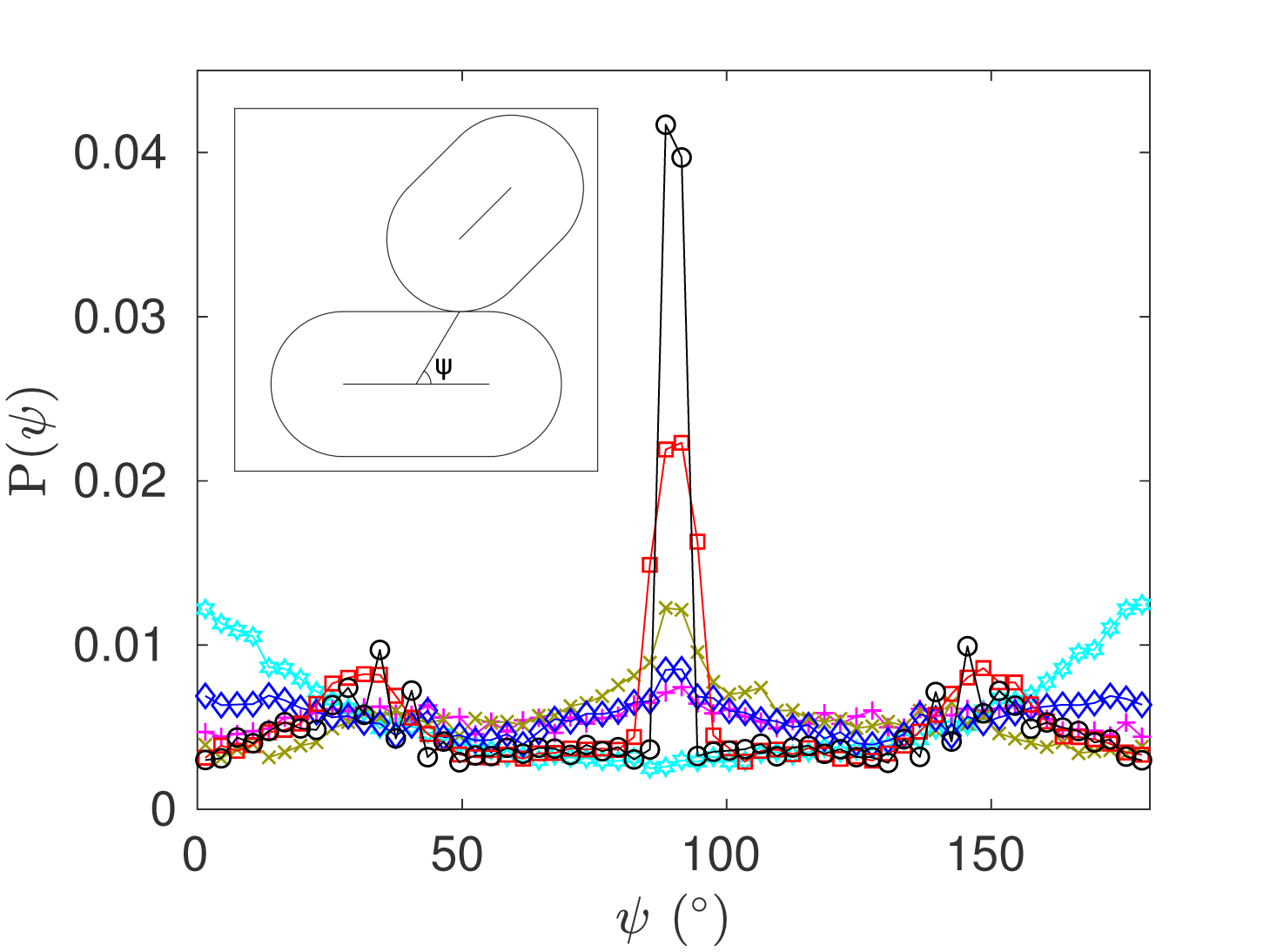}
\caption{Probability distribution $P(\psi)$ of the contact angles $\psi$
in bidisperse ellipse ($N=480$) and circulo-line ($N=100$) packings at 
several asphericities:
ellipses at ${\cal A}-1=0.19$ (six-pointed stars), $5 \times
10^{-3}$ (plus signs), and $4 \times 10^{-5}$ (exes), and circulo-lines
at ${\cal A}-1=0.18$ (diamonds), $4 \times 10^{-3}$ (squares),
and $4\times 10^{-5}$ (circles). In the inset, we define the contact
angle $\psi$ as the angle between the shaft of a circulo-line and
the vector pointing from its center to the point of contact with 
another circulo-line. We use
a similar definition for the contact angle for ellipses. Packings of circulo-lines, as well
ellipses, favor parallel contacts, even at
small asphericities.}
\label{fig:angledistros}
\end{figure}

The above observations about the curvatures of the inequality
constraints in configuration space can help explain the distribution
of contact angles $P(\psi)$ in static packings of elongated
particles~\cite{tian,unpub} shown in Fig.~\ref{fig:angledistros}. This
figure shows that, even for packings of circulo-lines at very small
asphericities, parallel contacts are highly probable, despite the fact
that the range of angles for parallel contacts at low asphericities is
small. This behavior for $P(\psi)$ can be explained by the fact that
end-middle and parallel contacts can contribute to making a hypostatic
packing mechanically stable, whereas end-end contacts cannot. (See
Table~\ref{tab:example}.)  Thus, end-middle and parallel contacts
(whose contact angles are close to $90^{\circ}$ at low asphericities) must be present to
stabilize hypostatic packings of low-asphericity
circulo-lines.  As shown in Fig.~\ref{fig:angledistros},
$P(\psi)$ is similar for both ellipse and circulo-line packings.

\section{Conclusions and Future Directions}
\label{conclusion}

In this article, we carried out computational studies of static
packings of frictionless nonspherical particles in 2D.  We developed
an interparticle potential for circulo-lines and -polygons that
generates continuous pair forces and torques as a function of the
particle coordinates. As a result, we are able to compare the
structural and mechanical properties of mechanically stable packings
of nine different nonspherical particle shapes: circulo-lines,
-triangles, -pentagons, -octagons, -decagons, asymmetric dimers,
dumbbells, Reuleaux triangles, and ellipses. Our studies place a
particular emphasis on the question of which particle shapes give rise
to {\it hypostatic} mechanically stable packings with fewer contacts than
the isostatic number.

We conjecture that to form hypostatic mechanically stable packings,
frictionless, convex particles must satisfy the following two
criteria: (i) the particle has one or more nontrivial rotational
degrees of freedom, and (ii) the particle cannot be defined as a union
of a finite number of complete disks without changing its accessible
contact surface. If the particle does not satisfy both criteria, we
expect it to form isostatic packings. Packings of the nine particle
shapes we considered are consistent with this conjecture. Future
research can investigate methods to analytically prove this
conjecture~\cite{deepiso}.

We then studied the packing fraction $\phi$ and coordination number
$z$ at jamming onset for packings of a number of different types of
nonspherical shapes in 2D as a function of asphericity ${\cal A}$.  To
do this, we resolved the ambiguity in the constraint counting of
nearly parallel contacts of circulo-lines and -polygons using the
branched structure of the eigenvalue spectra of the dynamical matrix.
In future research, we will study the coordination number of packings
of sphero-cylinders and -polygons in 3D, and compare the results to those in
2D, since it is extremely unlikely for sphero-cylinders and -polygons
to form nearly parallel contacts.

We find that the packing fraction and coordination number obey
approximate master curves when plotted versus the asphericity.
Further, the eigenvalue spectra for different particle shapes, at the
same ${\cal A}$, collapse. These results suggest that asphericity is a
key parameter in determining the structure, geometry, and mechanical
properties of hypostatic packings.  For $n$-sided circulo-polygons,
there are $2n-3$ parameters that specify their shape. In future
studies, we will investigate additional shape parameters, such as the
ratios of the area moments and others~\cite{turk}, to better
understand the coupling between the shape parameter space and the
properties of hypostatic packings of nonspherical particles.

We also demonstrated that hypostatic packings of circulo-lines (and by
analogy circulo-polygons) are mechanically stable by showing that even
though the eigenvalues of the dynamical matrix for the quartic modes
tend to zero at zero pressure, the fourth derivatives of the total
potential energy in the directions of the quartic modes do not. Thus,
hypostatic packings of nonspherical particles are stable to
perturbations in all directions in configuration space.  Perturbations
in some directions give rise to quadratic potentials, whereas other
directions give rise to quartic potentials.  In the directions with 
quartic potentials, we expect large anharmonic contributions to the 
vibrational and mechanical response~\cite{hertzian}. 

In addition, we measured the curvatures of the inequality constraints
that arise from interparticle contacts in hypostatic packings of
circulo-lines to better understand the grain-scale mechanisms that
allow hypostatic packings to be mechanically stable.  The contacts in
isostatic disk packings give rise to inequality constraints with only
concave (negative) curvatures.  In contrast, hypostatic packings of
circulo-lines (and other nonspherical particles) possess different
types of contacts ({\it e.g.} end-end and end-middle). Some types
yield inequality constraints with concave curvatures and others yield
inequality constraints with convex curvatures.  We find that contacts
with convex inequality constraints are present even at small
asphericities. The contacts with convex inequality
constraints allow the feasible region of slightly undercompressed
hypostatic packings to be compact, bounded, and shrink to zero in the
limit that the free volume tends to zero.

\section*{Acknowledgments}

The authors acknowledge financial support from NSF Grant
Nos. CMMI-1462439 (C.O.), CMMI-1463455 (M.S.), and CBET-1605178
(C.O. and K.V.), NIH Training Grant, Grant No. 1T32EB019941 (K.V.),
and the Raymond and Beverly Sackler Institute for Biological,
Physical, and Engineering Sciences (C. O. and K. V.).  We also
acknowledge the China Scholarship Council that supported Weiwei Jin's
visit to Yale University. In addition, this work was supported by the High
Performance Computing facilities operated by, and the staff of, the
Yale Center for Research Computing. We thank T. Marschall and S. 
Teitel for helpful conversations.

\appendix

\section{Continuous Potential between Circulo-lines and -Polygons}
\label{app:potential}

The repulsive potential between two circulo-lines is given by
Eq.~\eqref{potential}, where $r_{ij}$ is the magnitude of ${\vec
  r}_{ij}$, which points from the location where the force is applied on
circulo-line $j$ to the location  where the force is applied on
circulo-line $i$. These points of contact can be located on the ends or 
the shaft (middle) of a circulo-line. In this Appendix, we define the
overlap distance $\delta = \sigma_{ij} - r_{ij}$, which will depend 
on the type of contact that occurs between two circulo-lines. 

\subsection{Types of Contacts}
\label{sec:contacts}

There are three types of interparticle contacts that occur in packings
of circulo-lines: 1) the end of one circulo-line is in contact with
the middle of another (Fig.~\ref{fig:endmid}), 2) the shafts of two
circulo-lines are in contact and the circulo-lines are nearly parallel
(Figs.~\ref{fig:parallel} and~\ref{fig:nearparallel}), and 3) the ends
of two circulo-lines are in contact (Fig.~\ref{fig:endend}).  Below, 
we define the overlap distance $\delta$ in the circulo-line potential 
(Eq.~\ref{potential}) for each type of contact.   

\subsubsection{End-middle Contacts}
\label{sec:endmid}

\begin{figure}
\centering
\includegraphics[width=0.5\textwidth]{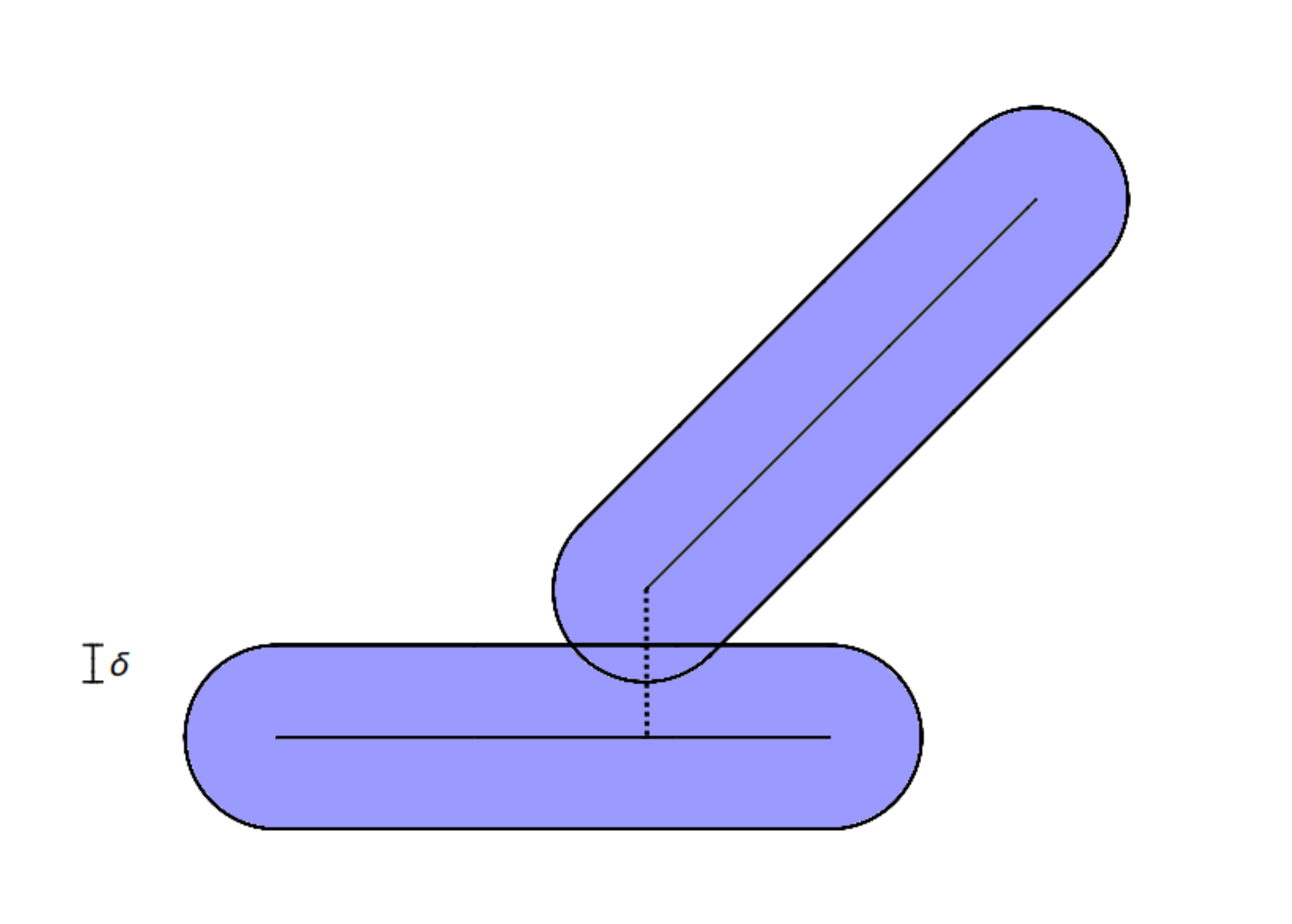}
\caption{A contact between the endcap of one circulo-line and the 
middle of another. The separation between the circulo-lines $r_{ij}$ is 
indicated by the dotted line
between the circulo-line shafts.  The separation vector ${\vec r}_{ij}$ 
connects the end of the shaft of one circulo-line to the shaft of the 
other and is perpendicular to shaft of the circulo-line with the 
middle contact.  The overlap is defined as $\delta=\sigma_{ij}-r_{ij}$, where $\sigma_{ij}$ is the sum of the endcap radius and the half-width of the 
shaft it overlaps.}
\label{fig:endmid}
\end{figure}

End-middle contacts occur when the endcap of one circulo-line makes
contact with the middle of another circulo-line, but does not overlap
with either of the other circulo-line's endcaps.  (See
Fig.~\ref{fig:endmid}.) In this case, we assume that the separation
vector ${\vec r}_{ij}$ between circulo-lines points from the end of
the shaft of the circulo-line with the end contact to the shaft of the
other circulo-line. ${\vec r}_{ij}$ is perpendicular to the shaft of
the circulo-line with the middle contact. The overlap between
circulo-lines with an end-middle contact is $\delta = \sigma_{ij} -
r_{ij}$, as shown in Fig.~\ref{fig:endmid}.

\subsubsection{Parallel and Nearly Parallel Contacts}
\label{sec:paracon}

\begin{figure}
\centering
\includegraphics[width=0.5\textwidth]{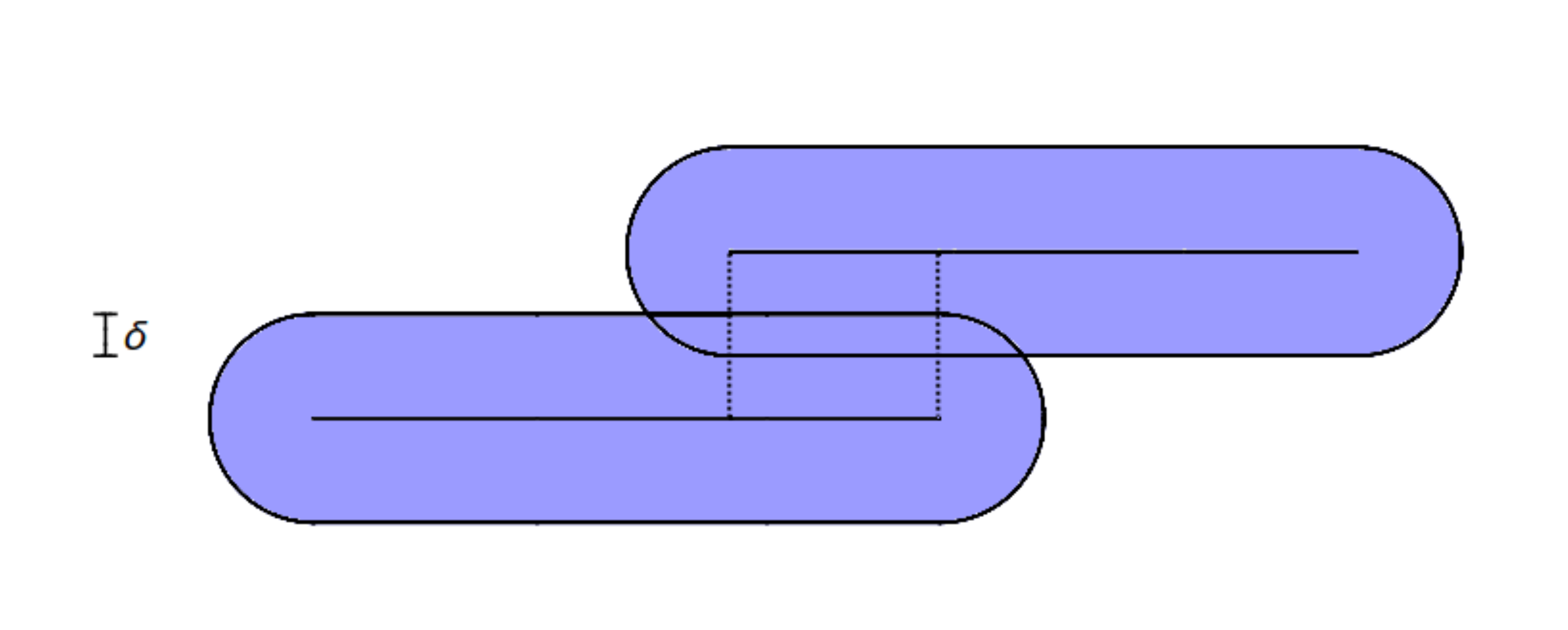}
\caption{When two circulo-lines possess a parallel contact, it is counted 
twice, as two end-middle contacts. The separations 
$r_{ij}$ are depicted by the vertical dotted lines between the 
circulo-line shafts, and ${\vec r}_{ij}$ points from 
the end of one of the shafts to the other and is perpendicular to the 
shafts. The overlap is given by $\delta = \sigma_{ij}-r_{ij}$ for 
both contacts, where $\sigma_{ij}$ is the sum of the endcap radius 
and the half-width of the shaft it overlaps.
}
\label{fig:parallel} 
\end{figure}

\begin{figure}
\centering
\includegraphics[width=0.5\textwidth]{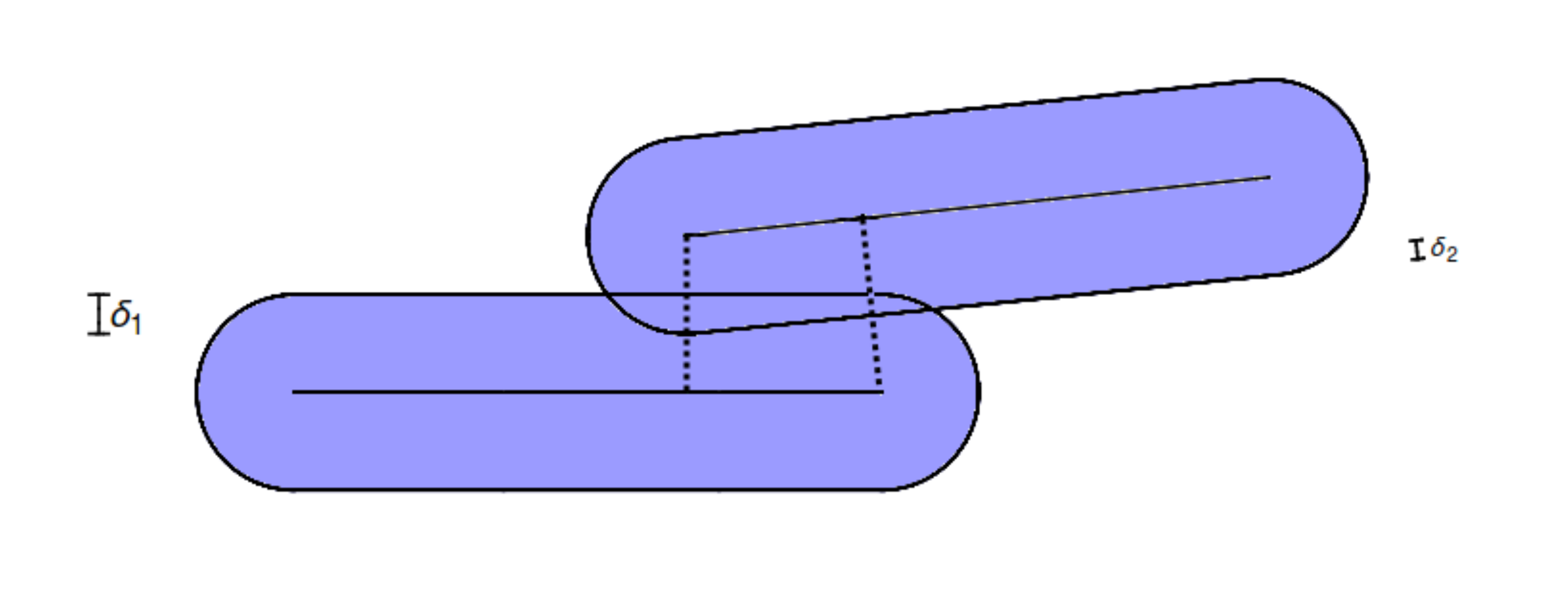}
\caption{When the shafts of two contacting circulo-lines are 
close to parallel, such 
that both endcaps overlap the shaft of the other circulo-line, the 
contact is counted twice as for parallel contacts. Each separation 
vector ${\vec r}_{ij}$ (depicted by dotted lines
between circulo-line shafts) points from the end of the shaft of 
the circulo-line with an end contact to the shaft of the circulo-line 
with the middle contact, and is perpendicular to the shaft of the circulo-line
with the middle contact. The overlaps are given by $\delta_{1,2} = 
\sigma_{ij}-r_{ij}$ for the two contacts with different separations, where $\sigma_{ij}$ is the sum of the endcap radius and the half-width of the 
shaft it overlaps.}
\label{fig:nearparallel}
\end{figure}

For parallel and nearly parallel contacts, an endcap of both  circulo-lines 
overlaps the shaft of the other circulo-line. In this case, the spring 
potential in Eq.~\ref{potential} for both overlaps is calculated as 
for end-middle contacts. If the circulo-lines are parallel, as in
Fig.~\ref{fig:parallel}, $\delta =\sigma_{ij} - r_{ij}$ is the same for both
overlaps. However, for nearly parallel contacts, as in
Fig.~\ref{fig:nearparallel}, the separations are different for the two 
end-middle overlaps. This method for treating end-middle, parallel, and nearly 
parallel contacts ensures continuity of the potential as a function of 
the particle coordinates. If the circulo-lines in Fig.~\ref{fig:nearparallel}
rotate until their orientations match Fig.~\ref{fig:endmid}, the
potential, force, and torque must all change continuously. Using our 
method, $\delta_2$ decreases continuously to zero as the contact evolves 
from that in Fig.~\ref{fig:nearparallel} to that in Fig.~\ref{fig:endmid}. 
In addition, $\delta_1$ decreases continuously to zero as the circulo-lines in
Fig.~\ref{fig:nearparallel} rotate until $\delta_2$ is the only
overlap. 

\subsubsection{End-End Contact}
\label{sec:endend}

\begin{figure}
\centering
\includegraphics[width=0.5\textwidth]{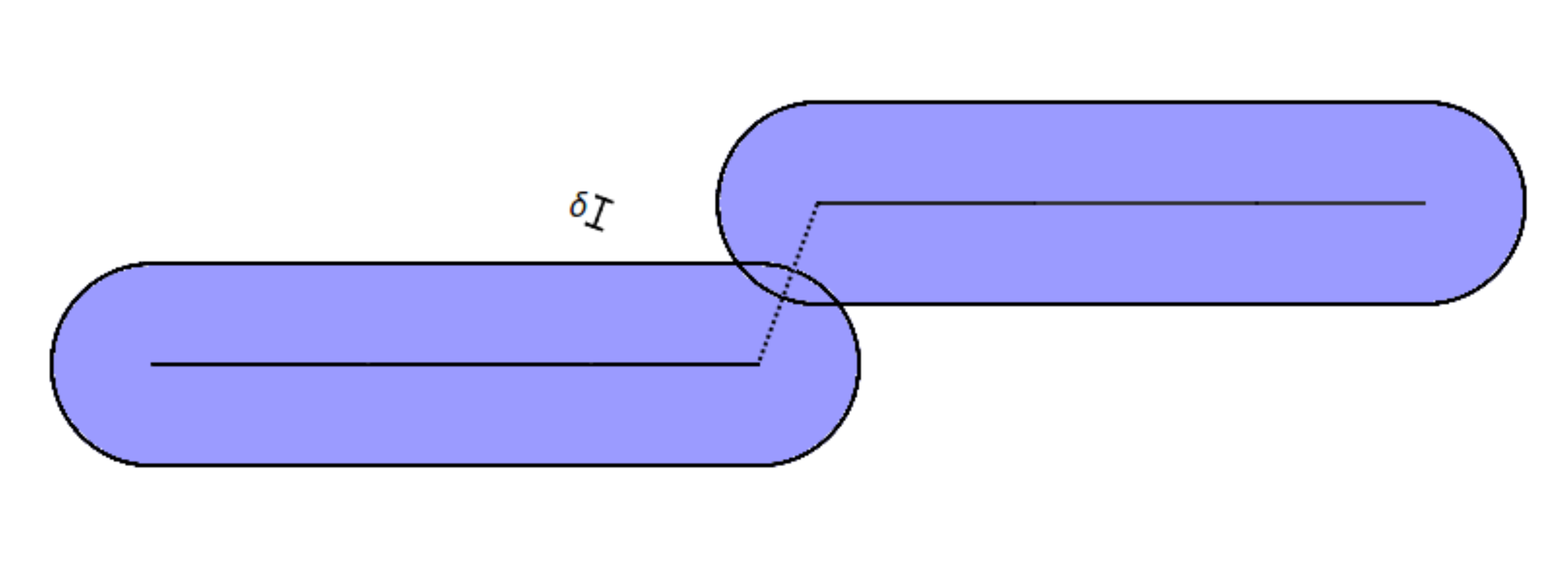}
\caption{An end-end contact between the endcaps of two circulo-lines. 
The separation ${\vec r}_{ij}$ between circulo-lines
(dotted line) gives the distances between the ends of the shafts 
of the two circulo-lines. The overlap is given by $\delta=
\sigma_{ij}-r_{ij}$, where $\sigma_{ij}$ is the sum of the radii 
of the endcaps.}
\label{fig:endend}
\end{figure}

Also, suppose that we slide the two circulo-lines in
Fig.~\ref{fig:parallel} away from each other until they are similar to
the configuration in Fig.~\ref{fig:endend} and form an end-end 
contact.  In this case, we assume that the two
overlap potentials add together as soon as the two relevant ends of the 
circulo-line shafts
slide past each other. Therefore, to ensure continuity, the interaction 
potential for an end-end
contact must be twice as large as that for an end-middle
contact. Hence, we use $U=k\delta^2$ for end-end contacts.

\begin{figure}
\centering
\includegraphics[width=0.5\textwidth]{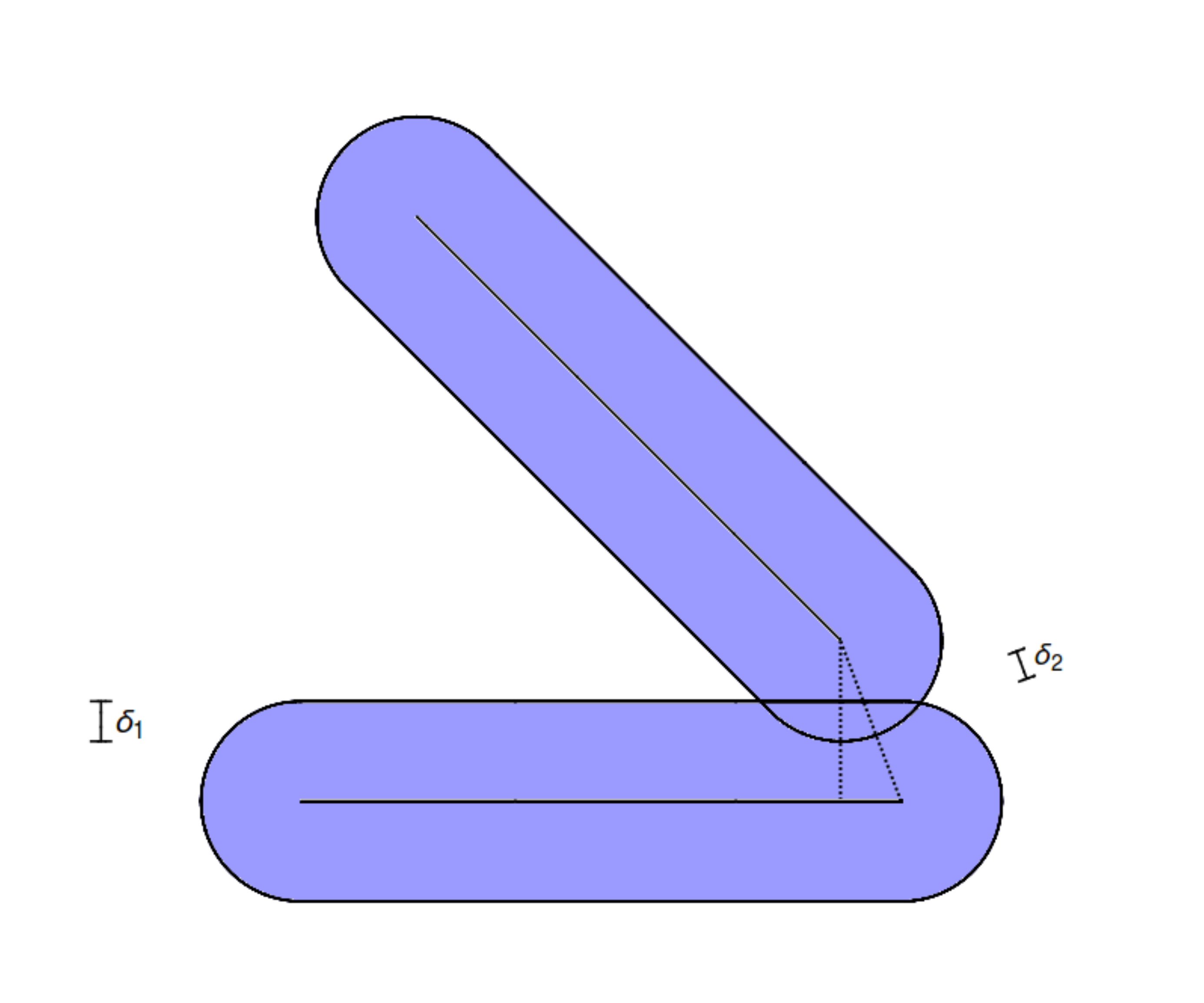}
\caption{When a contact between circulo-lines is close to the boundary 
between the end of one and
middle of another circulo-line, we include both the end-end and
end-middle overlaps to ensure continuity as the contact crosses
the boundary. The separation vector ${\vec r}_{ij}$ associated 
with the end-middle overlap, $\delta_1$ (vertical dotted line), points
from the end of the shaft of one circulo-line to the shaft of the other, 
such that it is perpendicular to the shaft. The separation vector associated 
with the end-end overlap, $\delta_2$ (diagonal dotted
line), points from the end of the shaft of one of the circulo-lines 
to the end of the shaft of the other overlapping circulo-line.}
\label{fig:eeaem}
\end{figure}

\begin{figure}
\centering
\includegraphics[width=0.5\textwidth]{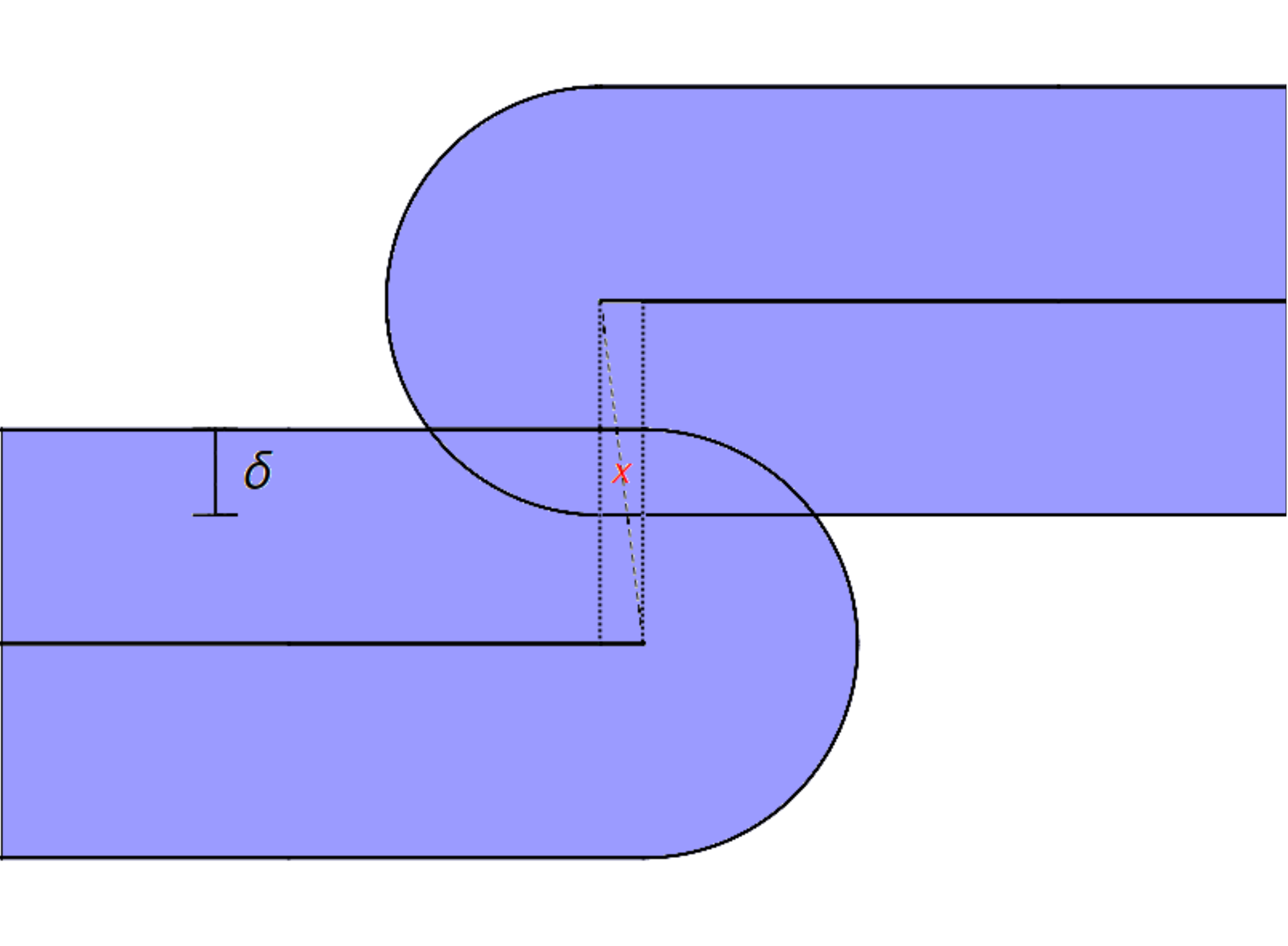}
\caption{We do not calculate the end-end overlap potential for parallel
and nearly parallel circulo-lines since this would lead to a 
discontinuity if the configuration transitions to an end-end contact by
sliding. Instead, we choose the circulo-line separations for 
parallel and nearly parallel contacts as 
shown in Fig.~\ref{fig:parallel}. The separation associated with 
an end-end contact is indicated by the dashed line with an $x$ in the middle.}
\label{fig:parallelnoee}\end{figure}

However, this treatment of end-end contacts creates a discontinuity
for the configuration in Fig.~\ref{fig:eeaem}. If we imagine sliding
the circulo-lines past each other until the overlap $\delta_1$ is
associated with an end-end contact, the potential will suffer a
discontinuous jump from $\tfrac{1}{2}k\delta_1^2$ to $k\delta_1^2$
since the end-end contact potential is twice as large as an end-middle
potential. To remedy this discontinuity, we add the end-end contact
potential between the two relevant endpoints as soon as they become
close enough to overlap. However, we do not make the end-end potential
twice as large in this case. Hence, the potential in this case is
given by $U=\tfrac{1}{2}k\left(\delta_1^2+\delta_2^2\right)$. Thus,
when we perform that same sliding transformation, the potential will
grow continuously from $\tfrac{1}{2}k\delta_1^2$ to $k\delta_1^2$ as
$\delta_2$ grows continuously from 0 to $\delta_1$.
Note that we do not add this end-end overlap potential if two
end-middle contacts are present, as in Fig.~\ref{fig:parallelnoee},
because in that case, the potential will already change continuously
as described in the previous subsection, and hence there is no
discontinuity to remedy.

\subsection{Generalizing the Circulo-line Potential to Circulo-Polygons}
\label{sec:genpoly}

\begin{figure}
\centering
\includegraphics[width=0.5\textwidth]{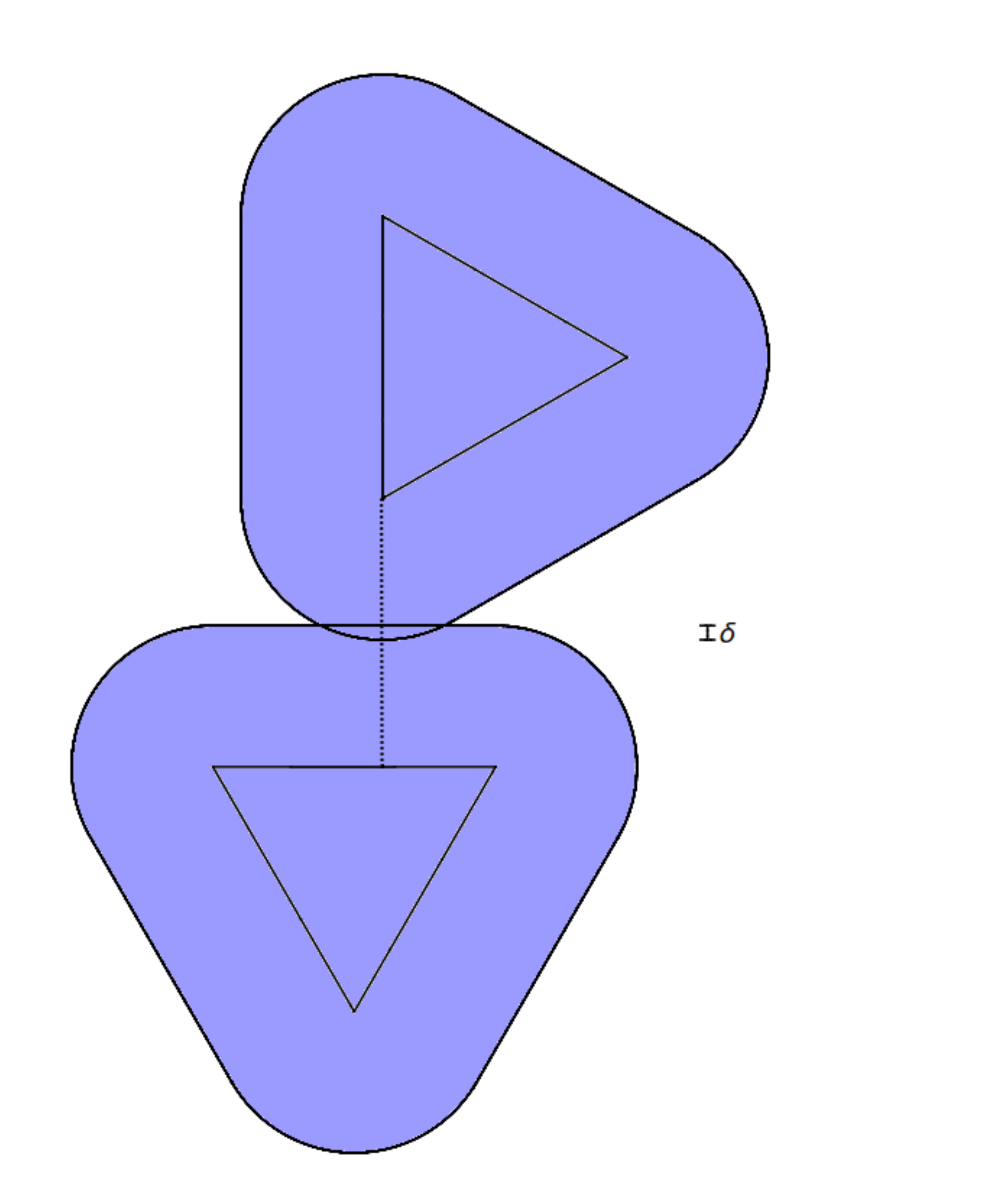}
\caption{A schematic of an end-middle contact between two circulo-triangles.  
The separation $r_{ij}$ between circulo-triangles (dotted line) 
is given by the perpendicular distance between the vertex of the circulo-triangle with the end contact and the side of the closest side of the triangle 
in the circulo-triangle with the middle contact. The overlap $\delta=\sigma_{ij} - r_{ij}$, where $\sigma_{ij}$ is the sum of the endcap radius and 
half-width of the shaft it overlaps.}
\label{fig:triendmid}
\end{figure}

Generalizing our continuous circulo-line potential to circulo-polygons,
such as those pictured in Fig.~\ref{fig:triendmid}, is
straightforward. We simply calculate the potential between all pairs
of circulo-lines that comprise each circulo-polygon. For example, in
Fig.~\ref{fig:triendmid}, since the vertex of the top circulo-triangle
is shared by two circulo-lines, we count the
end-middle contact twice, and hence the overlap potential is
$U=k\delta^2$.

\section{Generation of Circulo-Polygons}
\label{app:polyshape}

\begin{figure}[!htpb]
\centering
\includegraphics[width=0.5\textwidth]{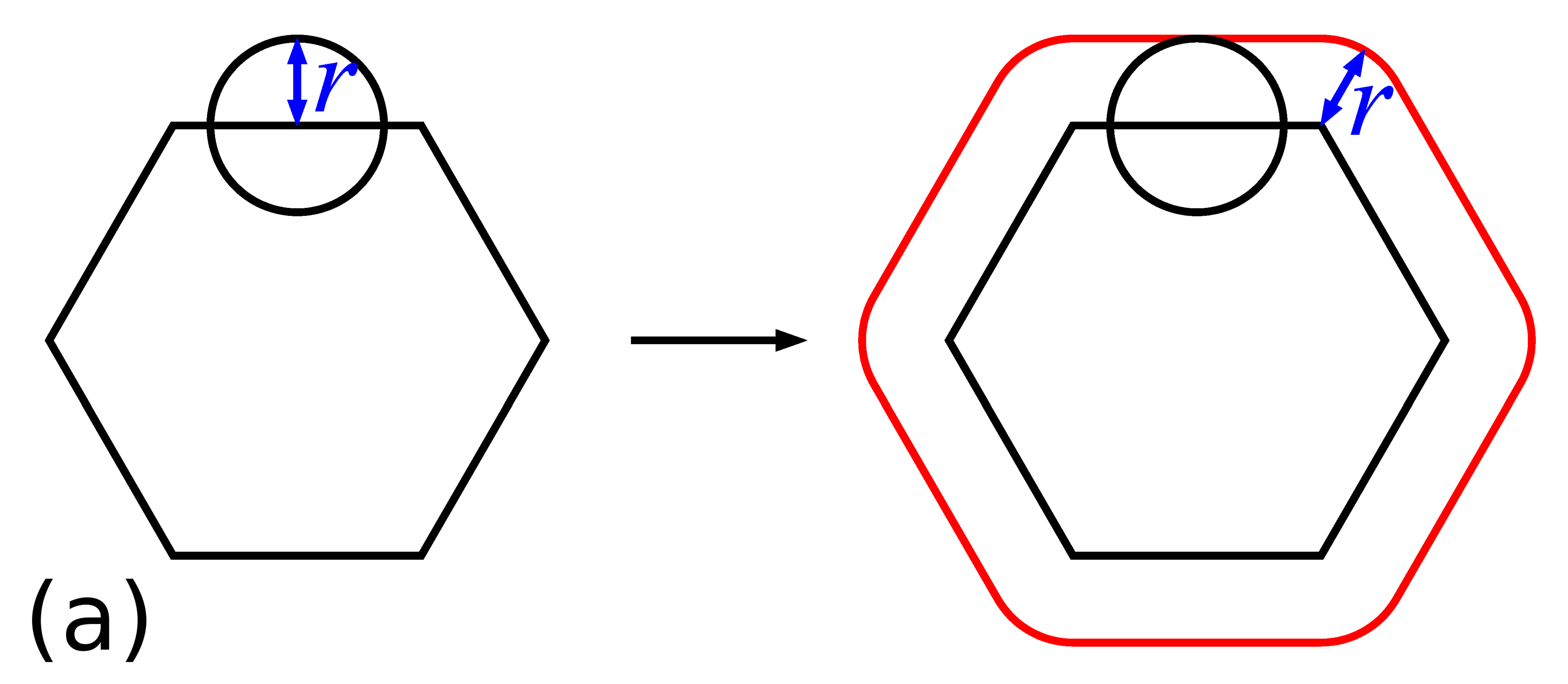}
\includegraphics[width=0.25\textwidth]{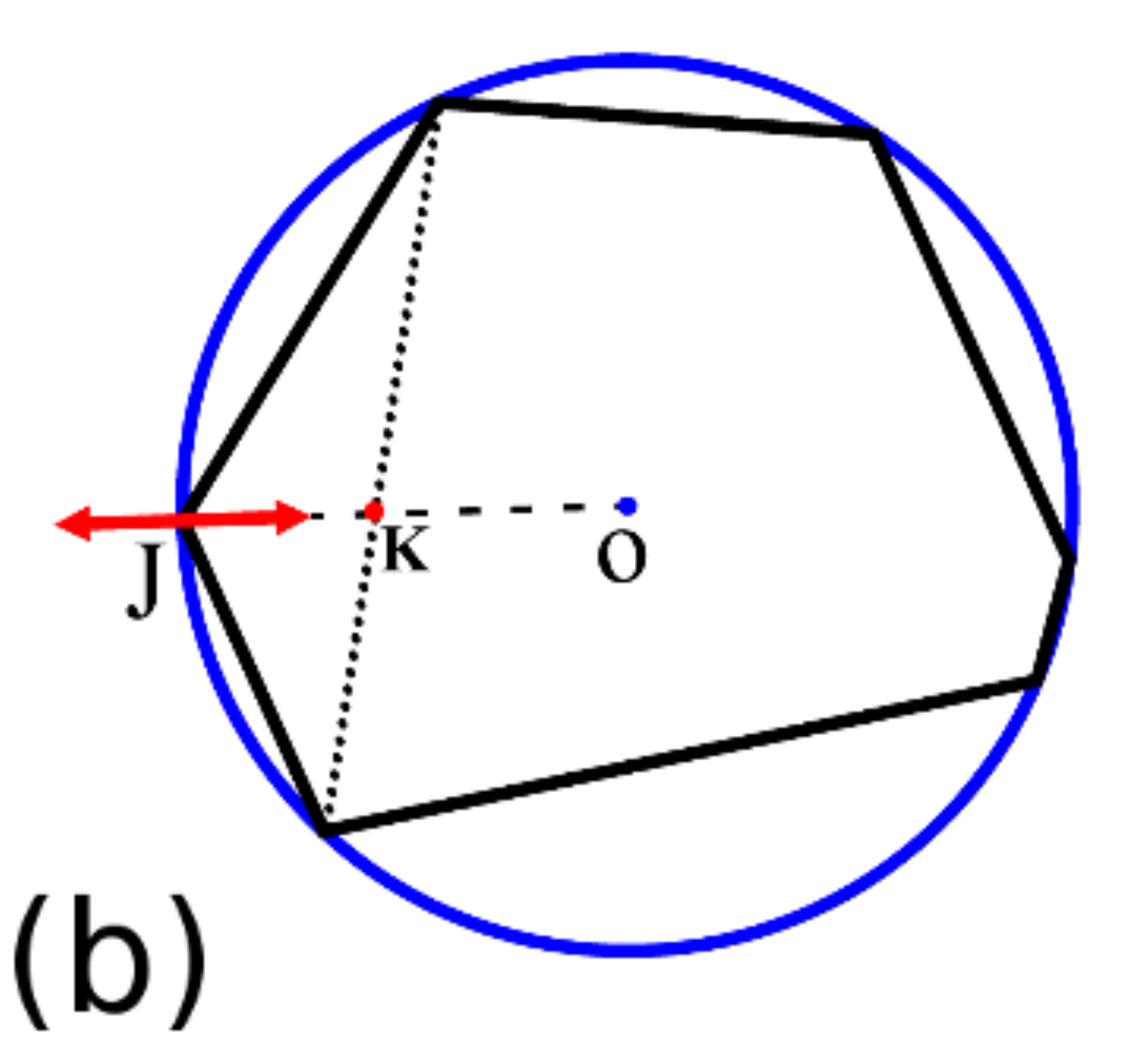}
\caption{(a) A circulo-polygon is the Minkowski sum of a polygon 
({\it e.g.} the regular hexagon on the left) and a disk with radius $r$, which is
``swept'' along the edges of the polygon to form the circulo-polygon
(the red shape on right). (b) Schematic of the generation of 
a circulo-polygon at a given asphericity ${\cal A}$. First, we 
select $n$ random
points on a unit circle (blue curve) with origin $O$ (where $n$ is the desired
number of edges). We then randomly choose vertex $J$
and either stretch or shorten the distance between $J$ and $O$
(dashed line) by an amount randomly chosen between 0 and the distance $JK$ to adjust ${\cal A}$ to match the target asphericity.}
\label{fig:circulopoly}
\end{figure}

A circulo-polygon is formed through a Minkowski sum of a polygon and a
disk with a radius $r$~\cite{minkowski}, which is equivalent to the
sweeping of the disk around the profile of the polygon as in
Fig.~\ref{fig:circulopoly} (a).  The shape of a circulo-polygon with
$n$ edges is fully specified by $2n-3$ independent parameters.  In
this work, we focus on the asphericity shape parameter ${\cal A}$,
which measures the deviation of a given shape from a circle in 2D.

We study bidisperse packings of circulo-polygons with asphericity
${\cal A}$ for which half of the circulo-polygons are large and half
are small. The large circulo-polygons have areas that satisfy $a_L =
1.4^2 a_S$, where $a_{L,S}$ is the area of the large and small
circulo-polygons, respectively. The large circulo-polygons (and small
ones) have different shapes at the same ${\cal A}$.  We generate
different circulo-polygons at the same ${\cal A}$ using the following
two-step approach: 1) We first randomly select $n$ points on a unit disk as the
vertices of an $n$-sided polygon. The radius $r$ of the
circulo-polygon is set to be $n$ percent of the perimeter of the
polygon. 2) If the asphericity of the current circulo-polygon is smaller
than the target $\cal A$, a vertex $J$ is randomly chosen and then
stretched or shortened along the direction between the vertex $J$ and
the center $O$ of the unit disk, by a distance randomly chosen between 0
and the distance between $J$ and the intersection of $JO$ with the line segment
connecting the two neighboring vertices, as shown in
Fig.~\ref{fig:circulopoly} (b). This deformation is accepted only if
the asphericity of the new shape is closer to $\cal A$ than the
original and the new shape is still convex. If the asphericity of the
current circulo-polygon exceeds ${\cal A}$, the radius $r$
is increased to match the target $\cal A$. We repeat step $2$ until a 
circulo-polygon with ${\cal A}$ is obtained.

\section{Dynamical Matrix Elements of Circulo-Polygon Packings}
\label{app:dynmat}

In this Appendix, we provide explicit expressions for the dynamical
matrix elements for static packings of circulo-polygons that interact
via the purely repulsive linear spring potential in
Eq.~\ref{potential}. In this expression, $R_i$ is radius that forms
the edge of circulo-polygon $i$ and ${\vec r}_{ji}$ is the
separation vector from circulo-polygon $i$ to $j$, which is
is given by
\begin{equation}
\vec{r}_{ji} = \vec {c}_{ji} + {\cal R}_j\vec {v}_n - {\cal R}_i\vec {v}_m + {\cal R}_j\hat {u}_n l_n - {\cal R}_i\hat {u}_m l_m,
\end{equation}
where $\vec{c}_{ji}=(x_j-x_i, y_j-y_i) \equiv (p, q)$ is the
center-to-center separation between circulo-polygons, ${\cal R}_i =
\left[\begin{array}{cccc}\cos\theta_i & -\sin\theta_i\\\sin\theta_i &
    \cos\theta_i\end{array}\right]$ is the rotation matrix in 2D,
$\theta_i$ is the orientation of particle $i$ relative to the
$x$-axis, $\vec {v}_m$ is the vector from the center of particle $i$
to the center of its corresponding edge $m$ when particle $i$ at zero
rotation, ${\hat u}_m$ is the unit vector along edge $m$ at zero
rotation, and $l_m$ indicates the distance between the contact point
and the center of edge $m$.  

The dynamical matrix requires the calculation of the second
derivatives of the total potential energy $U$, and can be expressed in
terms of the first and second derivatives of the contact distance
$r_{ji}$ with respect to the particle coordinates:
\begin{equation}
\frac{\partial ^2 U}{\partial \xi_i \partial \xi_j} = \frac{\partial {r_{ji}}}{\partial \xi_i}\frac{\partial {r_{ji}}}{\partial \xi_j} - \left(\sigma_{ji} - r_{ji}\right)\frac{\partial ^2{r_{ji}}}{\partial \xi_i \partial \xi_j},
\end{equation}
where $\xi_i=x_i$, $y_i$, or $\theta_i$,
\begin{equation}
\frac{\partial {r_{ji}}}{\partial \xi_i} = \frac{\vec{r}_{ji}}{r_{ji}} \cdot \frac{\partial \vec{r}_{ji}}{\partial \xi_i},
\end{equation}
and
\begin{equation}
\frac{\partial ^2{r_{ji}}}{\partial \xi_i \partial \xi_j} = \frac{1}{r_{ji}} \left(\frac{\partial \vec{r}_{ji}}{\partial \xi_i} \cdot \frac{\partial \vec{r}_{ji}}{\partial \xi_j} + \vec{r}_{ji} \cdot \frac{\partial^2 \vec{r}_{ji}}{\partial \xi_i \partial \xi_j} - \frac{\partial {r_{ji}}}{\partial \xi_i}\frac{\partial {r_{ji}}}{\partial \xi_j}\right).
\end{equation}
There are two types of contacts among circulo-polygons. The
first type is a vertex-to-edge contact. Assuming the end point of edge
$n$ on particle $j$ is in contact with edge $m$ on particle $i$,
$l_n$ is half of the length of edge $n$, and $l_m$ can be written
as
\begin{equation}
l_m = \left(\vec c_{ji}+{\cal R}_j\vec{v}_n-{\cal R}_i\vec{v}_m+{\cal R}_j\hat{u}_n l_n\right) \cdot {\cal R}_i\vec{u}_m.
\label{ell}
\end{equation}
In Eq.~\ref{ell}, $\vec{v}_m$, $\vec{v}_n$, $\hat{u}_m$, and
$\hat{u}_n$ are defined as
\begin{equation}
\vec{v}_m = \left(m\cos a, m\sin a\right),
\end{equation}
\begin{equation}
\vec{v}_n = \left(n\cos b, n\sin b\right),
\end{equation}
\begin{equation}
\hat{u}_m = \left(\cos d, \sin d\right),
\end{equation}
and
\begin{equation}
\hat{u}_n = \left(\cos e, \sin e\right),
\end{equation}
where $m$ and $n$ are the magnitudes of $\vec{v}_m$ and $\vec{v}_n$,
respectively, and the angles $a$, $b$, $d$, and $e$ are the orientations of
$\vec{v}_m$, $\vec{v}_n$, $\hat{u}_m$, and $\hat{u}_n$,
respectively. The contact distance $r_{ji}$ is
\begin{multline}
r_{ji}=|q\cos(d+\theta_i)-m\sin(a-d)+n\sin\beta+ \\ l_n \sin\Delta-p\sin(d+\theta_i)|,
\end{multline}
where
\begin{equation}
\beta = b-d+\theta_j-\theta_i
\end{equation}
and
\begin{equation}
\Delta = e-d+\theta_j-\theta_i.
\end{equation}

The nonzero first and second derivatives can be expressed  as:
\begin{equation}
\label{begineq}
\frac{\partial {r_{ji}}}{\partial x_i} = \Xi\, \sin (d+\theta_i),
\end{equation}
\begin{equation}
\frac{\partial {r_{ji}}}{\partial y_i} = -\Xi\, \cos (d+\theta_i),
\end{equation}
\begin{multline}
\frac{\partial {r_{ji}}}{\partial \theta_i} = -\Xi\, \bigl[n\cos \beta+l_n\cos\Delta+ \\ p\cos(d+\theta_i)+q\sin(d+\theta_i)\bigr],
\end{multline}
\begin{equation}
\frac{\partial {r_{ji}}}{\partial x_j} = -\frac{\partial {r_{ji}}}{\partial x_i},
\end{equation}
\begin{equation}
\frac{\partial {r_{ji}}}{\partial y_j} = -\frac{\partial {r_{ji}}}{\partial y_i},
\end{equation}
\begin{equation}
\frac{\partial {r_{ji}}}{\partial \theta_j} = -\Xi \, \left[n\cos\beta+l_n\cos\Delta\right],
\end{equation}
\begin{equation}
\frac{\partial^2 {r_{ji}}}{\partial x_i\partial \theta_i} = -\frac{\partial {r_{ji}}}{\partial y_i},
\end{equation}
\begin{equation}
\frac{\partial^2 {r_{ji}}}{\partial y_i\partial \theta_i} = \frac{\partial {r_{ji}}}{\partial x_i},
\end{equation}
\begin{multline}
\frac{\partial^2 {r_{ji}}}{\partial \theta_i\partial \theta_i} = -\Xi \, \bigl[n\sin \beta+l_n\sin\Delta- \\ p\sin(d+\theta_i)+q\cos(d+\theta_i)\bigr],
\end{multline}
\begin{equation}
\frac{\partial^2 {r_{ji}}}{\partial x_j\partial \theta_i} = \frac{\partial {r_{ji}}}{\partial y_i},
\end{equation}
\begin{equation}
\frac{\partial^2 {r_{ji}}}{\partial y_j\partial \theta_i} = -\frac{\partial {r_{ji}}}{\partial x_i},
\end{equation}
\begin{equation}
\frac{\partial^2 {r_{ji}}}{\partial \theta_j\partial \theta_i} = \Xi \, \left[n\sin\beta+l_n\sin\Delta\right],
\end{equation}
and
\begin{equation}
\label{endeq}
\frac{\partial^2 {r_{ji}}}{\partial \theta_j\partial \theta_j} =
-\frac{\partial^2 {r_{ji}}}{\partial \theta_j\partial \theta_i}.
\end{equation}
In the expressions in Eqs.~\ref{begineq}-\ref{endeq}, $\Xi$ is defined as:
\begin{multline}
\Xi=\textrm{Sgn}\bigl(q\cos(d+\theta_i)-m\sin(a-d)+n\sin\beta+ \\ l_n \sin\Delta-p\sin(d+\theta_i)\bigr),
\end{multline}
where
\begin{equation}
\textrm{Sgn}\left(z\right)=
\begin{cases}
1, \quad &z>0 \\
0, \quad &z=0 \\
-1, \quad &z<0.
\end{cases}
\end{equation}
All of the other first and second derivatives are zero.

The second type of contact between circulo-polygons is a a contact
between two vertices. In this case, $l_m$ and $l_n$ are each half the
lengths of edges $m$ and $n$, respectively. The $x$- and $y$-components
of separation vector $\vec{r}_{ji}$ are
\begin{multline}
x_{ji} = p+n\cos(b+\theta_j)+l_n\cos(e+\theta_j)- \\ m\cos(a+\theta_i)-l_m\cos(d+\theta_i)
\end{multline}
and
\begin{multline}
y_{ji} = q+n\sin(b+\theta_j)+l_n\sin(e+\theta_j)- \\ m\sin(a+\theta_i)-l_m\sin(d+\theta_i).
\end{multline}

For vertex-vertex contacts, the nonzero first and second derivatives are:
\begin{equation}
\frac{\partial {r_{ji}}}{\partial x_i} = -\frac{x_{ji}}{r_{ji}},
\end{equation}
\begin{equation}
\frac{\partial {r_{ji}}}{\partial y_i} = -\frac{y_{ji}}{r_{ji}},
\end{equation}
\begin{multline}
\frac{\partial {r_{ji}}}{\partial \theta_i} = \frac{1}{r_{ji}}(-l_m\bigl[q\cos(d+\theta_i) -p\sin(d+\theta_i) + \\ n\sin\beta+l_n\sin\Delta\bigr]
+m\bigl[p\sin(a+\theta_i) -q\cos(a+\theta_i)- \\ n\sin\omega +l_n\sin\mu\bigr]),
\end{multline}
\begin{equation}
\frac{\partial {r_{ji}}}{\partial x_j} = -\frac{\partial {r_{ji}}}{\partial x_i},
\end{equation}
\begin{equation}
\frac{\partial {r_{ji}}}{\partial y_j} = -\frac{\partial {r_{ji}}}{\partial y_i},
\end{equation}
\begin{multline}
\frac{\partial {r_{ji}}}{\partial \theta_j} = \frac{1}{r_{ji}}(l_n\bigl[q\cos(e+\theta_j) -p\sin(e+\theta_j) - \\ m\sin\mu+l_m\sin\Delta\bigr]
+n\bigl[q\cos(b+\theta_j) -p\sin(b+\theta_j) + \\ m\sin\omega +l_m\sin\beta\bigr]),
\end{multline}
\begin{equation}
\frac{\partial \vec{r}_{ji}}{\partial x_i} \cdot \frac{\partial \vec{r}_{ji}}{\partial x_i} = \frac{\partial \vec{r}_{ji}}{\partial y_i} \cdot \frac{\partial \vec{r}_{ji}}{\partial y_i} = \frac{\partial \vec{r}_{ji}}{\partial x_j} \cdot \frac{\partial \vec{r}_{ji}}{\partial x_j} = \frac{\partial \vec{r}_{ji}}{\partial y_j} \cdot \frac{\partial \vec{r}_{ji}}{\partial y_j} =1,
\end{equation}
\begin{equation}
\frac{\partial \vec{r}_{ji}}{\partial x_i} \cdot \frac{\partial \vec{r}_{ji}}{\partial x_j} = \frac{\partial \vec{r}_{ji}}{\partial y_i} \cdot \frac{\partial \vec{r}_{ji}}{\partial y_j} = -1,
\end{equation}
\begin{equation}
\frac{\partial \vec{r}_{ji}}{\partial x_i} \cdot \frac{\partial \vec{r}_{ji}}{\partial \theta_i} = -m\sin(a+\theta_i)-l_m\sin(d+\theta_i),
\end{equation}
\begin{equation}
\frac{\partial \vec{r}_{ji}}{\partial x_i} \cdot \frac{\partial \vec{r}_{ji}}{\partial \theta_j} = n\sin(b+\theta_i)+l_n\sin(e+\theta_j),
\end{equation}
\begin{equation}
\frac{\partial \vec{r}_{ji}}{\partial y_i} \cdot \frac{\partial \vec{r}_{ji}}{\partial \theta_i} = m\cos(a+\theta_i)+l_m\cos(d+\theta_i),
\end{equation}
\begin{multline}
\\
\frac{\partial \vec{r}_{ji}}{\partial y_i} \cdot \frac{\partial \vec{r}_{ji}}{\partial \theta_j} = -n\cos(b+\theta_i)-l_n\cos(e+\theta_j),
\end{multline}
\begin{equation}
\frac{\partial \vec{r}_{ji}}{\partial \theta_i} \cdot \frac{\partial \vec{r}_{ji}}{\partial \theta_i} = l_m^2+m^2+2l_mm\cos(a-d),
\end{equation}
\begin{equation}
\frac{\partial \vec{r}_{ji}}{\partial \theta_i} \cdot \frac{\partial \vec{r}_{ji}}{\partial x_j} = -\frac{\partial \vec{r}_{ji}}{\partial x_i} \cdot \frac{\partial \vec{r}_{ji}}{\partial \theta_i},
\end{equation}
\begin{equation}
\frac{\partial \vec{r}_{ji}}{\partial \theta_i} \cdot \frac{\partial \vec{r}_{ji}}{\partial y_j} = -\frac{\partial \vec{r}_{ji}}{\partial y_i} \cdot \frac{\partial \vec{r}_{ji}}{\partial \theta_i},
\end{equation}
\begin{multline}
\frac{\partial \vec{r}_{ji}}{\partial \theta_i} \cdot \frac{\partial \vec{r}_{ji}}{\partial \theta_j} = -n(l_m\cos\beta+m\cos\omega)- \\ l_n(l_m\cos\Delta+m\cos\mu),
\end{multline}
\begin{equation}
\frac{\partial \vec{r}_{ji}}{\partial x_j} \cdot \frac{\partial \vec{r}_{ji}}{\partial \theta_j} = -\frac{\partial \vec{r}_{ji}}{\partial x_i} \cdot \frac{\partial \vec{r}_{ji}}{\partial \theta_j},
\end{equation}
\begin{equation}
\frac{\partial \vec{r}_{ji}}{\partial y_j} \cdot \frac{\partial \vec{r}_{ji}}{\partial \theta_j} = -\frac{\partial \vec{r}_{ji}}{\partial y_i} \cdot \frac{\partial \vec{r}_{ji}}{\partial \theta_j},
\end{equation}
\begin{equation}
\frac{\partial \vec{r}_{ji}}{\partial \theta_j} \cdot \frac{\partial \vec{r}_{ji}}{\partial \theta_j} = l_n^2+n^2+2l_nn\cos(b-e),
\end{equation}
and
\begin{multline}
\vec{r}_{ji} \cdot \frac{\partial^2 \vec{r}_{ji}}{\partial \theta_i\partial \theta_i} = -\frac{\partial \vec{r}_{ji}}{\partial \theta_i} \cdot \frac{\partial \vec{r}_{ji}}{\partial \theta_i} - \frac{\partial \vec{r}_{ji}}{\partial \theta_i} \cdot \frac{\partial \vec{r}_{ji}}{\partial \theta_j} + \\ p\frac{\partial \vec{r}_{ji}}{\partial y_i} \cdot \frac{\partial \vec{r}_{ji}}{\partial \theta_i}-q\frac{\partial \vec{r}_{ji}}{\partial x_i} \cdot \frac{\partial \vec{r}_{ji}}{\partial \theta_i}.
\end{multline}
\begin{multline}
\vec{r}_{ji} \cdot \frac{\partial^2 \vec{r}_{ji}}{\partial \theta_j\partial \theta_j} = -\frac{\partial \vec{r}_{ji}}{\partial \theta_j} \cdot \frac{\partial \vec{r}_{ji}}{\partial \theta_j} - \frac{\partial \vec{r}_{ji}}{\partial \theta_i} \cdot \frac{\partial \vec{r}_{ji}}{\partial \theta_j} + \\ p\frac{\partial \vec{r}_{ji}}{\partial y_i} \cdot \frac{\partial \vec{r}_{ji}}{\partial \theta_j}-q\frac{\partial \vec{r}_{ji}}{\partial x_i} \cdot \frac{\partial \vec{r}_{ji}}{\partial \theta_j},
\end{multline}
where 
\begin{equation}
\omega = b-a+\theta_j-\theta_i
\end{equation}
and
\begin{equation}
\mu = a-e-\theta_j+\theta_i.
\end{equation}
The other derivatives, $\frac{\partial
  \vec{r}_{ji}}{\partial \xi_i} \cdot \frac{\partial
  \vec{r}_{ji}}{\partial \xi_j}$ and $\vec{r}_{ji} \cdot \frac{\partial^2
  \vec{r}_{ji}}{\partial \xi_i\partial \xi_j}$, that are not listed  above 
are zero.

\bibliography{hypostatic_1.bib}

\end{document}